\documentclass[10pt,journal,letterpaper,compsoc]{IEEEtran}

\IEEEoverridecommandlockouts

\usepackage{multicol}

\usepackage{url}

\usepackage{algorithm}

\usepackage{epsfig}

\usepackage{cite}
\ifCLASSINFOpdf
   \usepackage{graphicx}
\else
\fi

\ifCLASSOPTIONcompsoc
\else
\fi

\usepackage[cmex10]{amsmath}
\usepackage{algorithmic}

\usepackage[font=footnotesize,caption=false]{subfig}
\usepackage[margin=0.5in]{geometry}

\begin{document}

\title{%
\tiny{This work has been submitted to the IEEE for possible publication. Copyright may be transferred without notice, after which this version may no longer be accessible.}\\
\Huge TVR -- Tall Vehicle Relaying in Vehicular Networks %
}

\author{
Mate Boban, %
Rui Meireles, %
Jo\~{a}o Barros, %
Peter Steenkiste, %
and Ozan K. Tonguz %
\IEEEcompsocitemizethanks{\IEEEcompsocthanksitem M. Boban is with the Department of Electrical and Computer Engineering, Carnegie Mellon University, USA. %
E-mail: mboban@cmu.edu.
\IEEEcompsocthanksitem R. Meireles is with the Department of Computer Science, Carnegie Mellon University, USA. He is also with Instituto de Telecomunica\c c\~oes, Faculdade de Engenharia da Universidade do Porto, Portugal. 
E-mail: rui@cmu.edu.
\IEEEcompsocthanksitem J. Barros is with Instituto de Telecomunica\c c\~oes, Faculdade de Engenharia da Universidade do Porto, Portugal. 
E-mail: jbarros@fe.up.pt.
\IEEEcompsocthanksitem P. Steenkiste is with the Departments of Computer Science and Electrical and Computer Engineering, Carnegie Mellon University, USA. 
E-mail: prs@cs.cmu.edu.
\IEEEcompsocthanksitem O. K. Tonguz is with the Department of Electrical and Computer Engineering, Carnegie Mellon University, USA. %
E-mail: tonguz@ece.cmu.edu.}

\thanks{This work was funded %
by the Portuguese Foundation for Science and Technology under the Carnegie Mellon $\mid$ Portugal program (grants SFRH/BD/33771/2009 and SFRH/BD/37698/2007) and the DRIVE-IN project (CMU-PT/NGN/0052/2008. http://drive-in.cmuportugal.org). }

}
\IEEEcompsoctitleabstractindextext{\begin{abstract}
Vehicle-to-Vehicle (V2V) communication is a core technology for enabling safety and non-safety applications in %
next generation Intelligent Transportation Systems. 
Due to relatively low heights of the antennas, %
V2V communication   
is often influenced by topographic features, man-made structures, and other vehicles located between the communicating vehicles. %
On highways, it was shown experimentally that vehicles can obstruct the line of sight (LOS) communication up to 50\% of the time; furthermore, a single obstructing vehicle can reduce the power at the receiver by more than 20~dB.
Based on both experimental measurements and simulations performed using a validated channel model, we show that the elevated position of antennas on tall vehicles improves communication performance. Tall vehicles can significantly increase the effective communication range, with an improvement of up to 50\% in certain scenarios. Using these findings, we propose a new V2V relaying scheme called Tall Vehicle Relaying (TVR) that takes advantage of better channel characteristics provided by tall vehicles. TVR distinguishes between tall and short vehicles and, where appropriate, chooses tall vehicles as next hop relays. We investigate TVR's system-level performance through a combination of link-level experiments and system-level simulations and show that it outperforms existing techniques.

\end{abstract}
}
\maketitle

\section{Introduction} \label{sec:Introduction}
A large number of Intelligent Transportation Systems (ITS) applications to be supported by vehicular ad-hoc networks (VANETs) rely on Vehicle-to-Vehicle (V2V) communication. These applications range from safety \cite{etsi09,chen05,boban12} to traffic management \cite{bai06,martinez09} and infotainment \cite{dikaiakos07,amadeo12}. The relatively low height of the antennas located on the vehicles makes V2V communication susceptible to line of sight (LOS) obstruction by non-communicating vehicles. The probability of having LOS communication decreases with distance, with less than a 50\% chance of LOS near the maximum V2V communication range~\cite{boban11}. Furthermore, the Dedicated Short-Range Communications (DSRC)~\cite{dsrc09} frequency band reserved for VANET communication is in the 5.9~GHz band. As noted by Parsons in~\cite{00parsons}, in this frequency band the ``propagation paths must have line-of-sight between the transmitting and receiving antennas, otherwise losses are extremely high''.
This has been empirically shown to be the case for V2V links in \cite{meireles10}, where a single large truck attenuated the received power between two passenger cars by 27~dB.  %
Consequently, obstructing vehicles cause a reduction of the effective communication range of up to 60\% and Packet Delivery Ratio (PDR) of up to 30\%, depending on the environment. %

Motivated by these findings, we explore how the adverse effects of vehicular obstructions can be ameliorated by opting for taller vehicles as next hop relays, whenever possible. %
We distinguish between tall vehicles, such as commercial and public transportation vehicles (vans, buses, trucks, etc.) 
and short vehicles (passenger cars). We base this distinction on the analysis performed in \cite{boban11}, which showed that the dimensions of the most popular passenger cars differ significantly from the dimensions of commercial freight and public transportation vehicles. Specifically, it was observed that the latter are, on average, more than 1.5~meters taller than personal vehicles. %
By separating the vehicles in this manner, we showed in~\cite{boban11_2} that the antennas mounted on top of tall vehicles experience a significantly better communication channel, 
which is not as affected by obstruction from other vehicles as is the case for 
short vehicles (i.e., the probability of having LOS conditions increases). 

This paper goes beyond the findings of~\cite{boban11_2} by: I) performing link-level experiments %
that provide insights into the end-to-end benefits of tall vehicle relaying; and II) introducing a tall vehicle relaying technique. %
Specifically, we use the one- and two-hop experiments to: a) quantify the benefits of tall vehicle relays in terms of received power and packet delivery ratio; and b) validate the channel model we use for the subsequent system-level simulation study of tall vehicle relaying. 
Additionally, %
we introduce the Tall Vehicle Relaying (TVR) technique,
a paradigm shift from the farthest relay technique, which selects the farthest tall vehicle in the direction of message destination. We compare the performance of TVR with two techniques: i) \emph{Farthest Neighbor}, which selects the farthest neighbor with which communication is possible; and ii) and \emph{Most New Neighbors}, which selects the vehicle with the largest number of new neighbors in the direction of message dissemination. %

The main contributions of this work can be summarized as follows:
\begin{itemize}
\item We perform real-world experiments to determine the benefits of using tall vehicles as relays; the results show that selecting tall vehicles is beneficial in terms of higher received power, smaller number of hops to reach the destination, %
and increased per-hop communication range (Section~\ref{sec:experiment});
\item We introduce the Tall Vehicle Relay (TVR) technique, which utilizes better channel characteristics provided by tall vehicles to make better relaying decisions (Section~\ref{sec:TVR});

\item We perform simulations to evaluate the system-level benefits of TVR. The results show that TVR matches existing techniques in low vehicle density scenarios and outperforms them in high density scenarios %
in terms of the number of hops needed to reach the destination, thus also decreasing the end-to-end delay (Section~\ref{sec:largeScale}).
\end{itemize}

The rest of the paper is organized as follows. %
Section~\ref{sec:motivation} motivates our study by quantifying the increase in proportion of line of sight links experienced by tall vehicles as compared to short vehicles. %
Section~\ref{sec:experiment} describes the results of the experimental study we performed to characterize the benefits of tall vehicles as relays. %
Section~\ref{sec:TVR} introduces the TVR technique, whereas Section~\ref{sec:largeScale} presents the results of the system-level simulations we performed to evaluate benefits of TVR.
Section~\ref{sec:RelatedWork} discusses related work. %
Finally, Section~\ref{sec:Discussion} %
concludes the paper.

\section{Tall Vehicles Increase the Chance of LOS Links}\label{sec:motivation}

The motivation for our study stems from the numerous previous studies (e.g., \cite{dhoutaut06,usdot06_2,Otto2009,meireles10,masui02}), which have shown %
that the resulting channel characteristics for LOS and non-LOS links are fundamentally different. To that end, in this section we analyze the effect of vehicle height on the probability of having LOS links. %

To assess the effect of tall vehicles on LOS communication, we require accurate information on vehicle positions and dimensions.
For accurate vehicle positioning, 
we leverage a dataset of real vehicle positions obtained from aerial photography of the A28 highway located near Porto, Portugal. The 404 vehicle dataset is described in Table~\ref{dataset}; more details on the dataset are available in~\cite{ferreira10}. In addition to vehicle location, this dataset specifies the heading and the length of each vehicle. %
To assign width and height %
to each vehicle, we used the empirically derived distributions of the dimensions of tall and short vehicles described in~\cite{boban11}. The heights of both types of vehicles are normally distributed, with a mean of 3.35 meters for tall and 1.5 meters for short vehicles. The standard deviation is 0.08 meters for both types.

\begin{table} 
	\centering
		\caption{\small Aerial photography dataset (A28 highway)}
		\begin{tabular}{|c c c c c|}
\hline \textbf{Highway} & \textbf{Length} & \textbf{\# Vehicles} & \textbf{\# Tall Vehicles} & \textbf{Veh. Density} \\
		 	\hline 
\hline A28 & 12.5~km & 404 & 58 (14.36\%) & 32.3~veh/km\\ 
\hline
\end{tabular}
	\label{dataset}
\end{table}

Using the aerial dataset, we determine: a) how often the LOS is blocked by non-communicating vehicles; and b) the difference in LOS blocking between short and tall vehicles.
For this purpose, we define the per-vehicle ratio of LOS links as follows.
For each vehicle, we determine the number of neighbors it has a LOS with, where a neighbor is any vehicle within a specified range around the observed vehicle. %
Then, we divide that number by the total number of neighbors. This gives the ratio of LOS links for a specific vehicle. By doing the same calculation for each vehicle and by separating the tall and short vehicles, we obtain the distribution of the ratio of LOS links. %

\begin{figure}
  \begin{center}
    \includegraphics[width=0.35\textwidth]{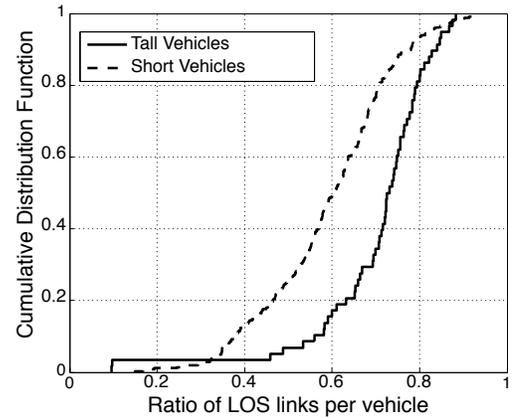}
     \caption{\small Cumulative Distribution Function of the per-vehicle ratio of LOS links within 750 meter range
     for tall and short vehicles based on the aerial photography dataset. %
}
      \label{PLOS-cars-trucks}
   \end{center}
\end{figure}

Figure~\ref{PLOS-cars-trucks} shows the difference in the ratio of LOS links for tall and short vehicles. %
The ratio of LOS links %
is notably higher for tall vehicles; 50\% of the short vehicles have more than 60\% of LOS links, whereas 
for tall vehicles, %
the value rises from 50\% to 80\%.

\section{Experimental Study Analyzing Benefits of Tall Vehicle Relays and Validating Channel Model Used for System-Level Simulations}\label{sec:experiment}
\begin{figure}
  \begin{center}
    \includegraphics[width=0.3\textwidth]{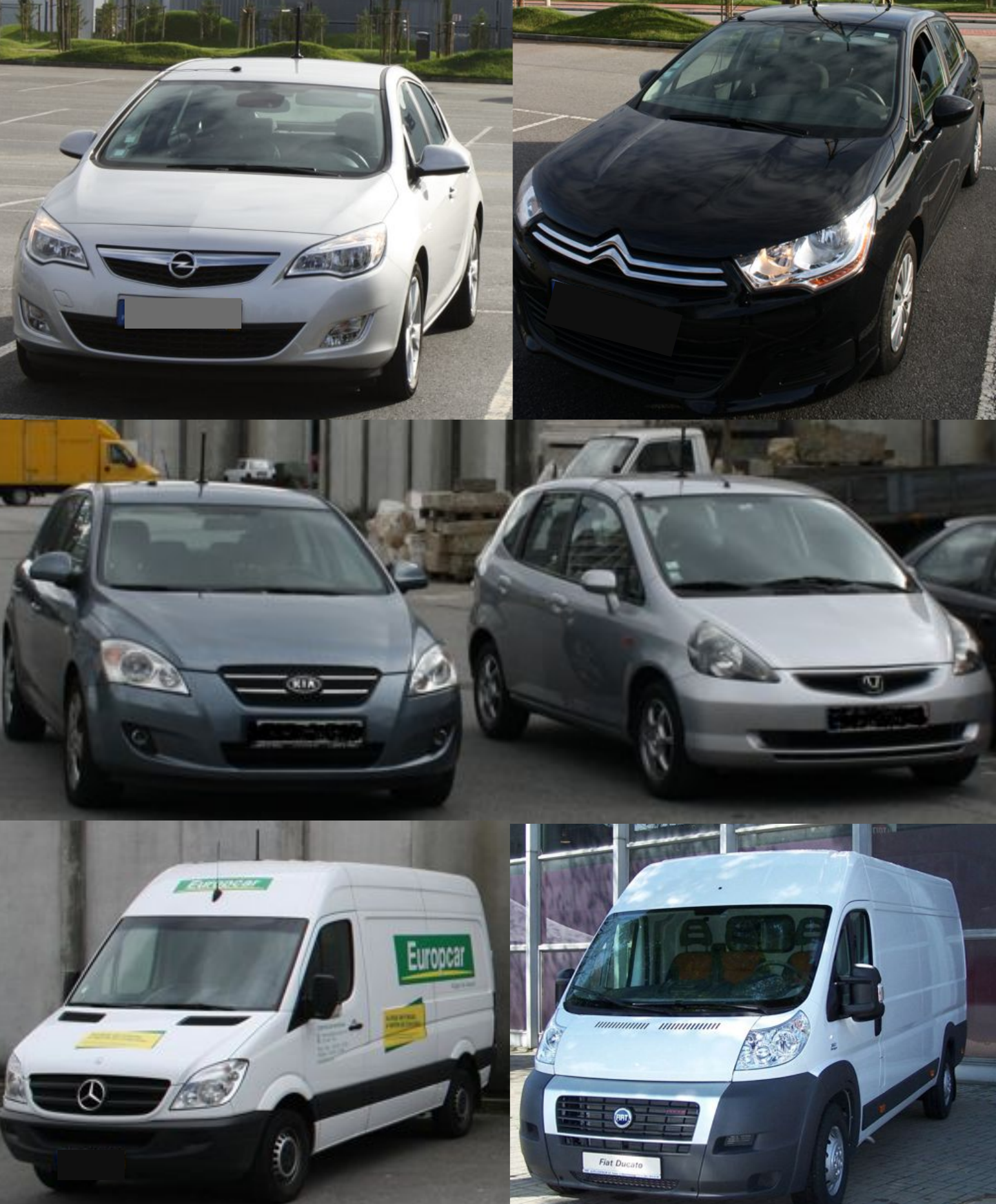}
     \caption{\small  Vehicles used in the experiments. Clockwise from top left: Opel Astra, Citroen C4, Honda Jazz, Fiat Ducato, Mercedes Sprinter, and Kia Cee'd. The four cars have a height of approximately 1.5 meters, which coincides with the statistical mean height for personal vehicles~\cite{boban11}, %
whereas both vans are approximately 2.5~meters tall.}
      \label{vehicles}
   \end{center}
\end{figure}

In this section, we describe the link-level measurements we performed to quantify the benefits of tall vehicle relays in terms of received power and packet delivery ratio. Additionally, we use the measurements to validate the channel model we employ for subsequent system-level simulation study of the benefits of tall vehicle relays (described in Section~\ref{sec:largeScale}). 
 Using regular passenger cars to represent the short vehicle class and full-size vans to represent the tall vehicle class, we performed experiments comprising two-node and three-node networks. Vehicles used in the experiments are depicted in Fig.~\ref{vehicles}; their dimensions are listed in Table~\ref{tab:dimensions}. %

\begin{table}
	\centering
		\caption{\small Dimensions of Vehicles Used in the Experiments} 
\begin{tabular}{|c c c c|} \hline
		 & \multicolumn{3}{c|}{\bf Dimensions (meters)} \\ %
		 \bf Vehicle & \bf Height & \bf Width & \bf Length \\ \hline \hline
		 \textbf{Passenger (short) vehicles}&&& \\ \hline
		2011 Citroen C4 & 1.491 & 1.789 & 4.329\\ \hline
		2011 Opel Astra & 1.510 & 1.814&  4.419\\ \hline
		2007 Kia Cee'd & 1.480 & 1.790 & 4.260\\ \hline
		2002 Honda Jazz & 1.525 & 1.676 & 3.845\\ \hline 
		\textbf{Commercial (tall) vehicles}&&& \\ \hline
		2010 Mercedes Sprinter & 2.591 & 1.989 & 6.680\\ \hline
		2010 Fiat Ducato & 2.524 & 2.025 & 5.943\\ \hline
		\end{tabular} 
\label{tab:dimensions}
\end{table}

\begin{figure}
  \begin{center}
    \includegraphics[width=0.4\textwidth]{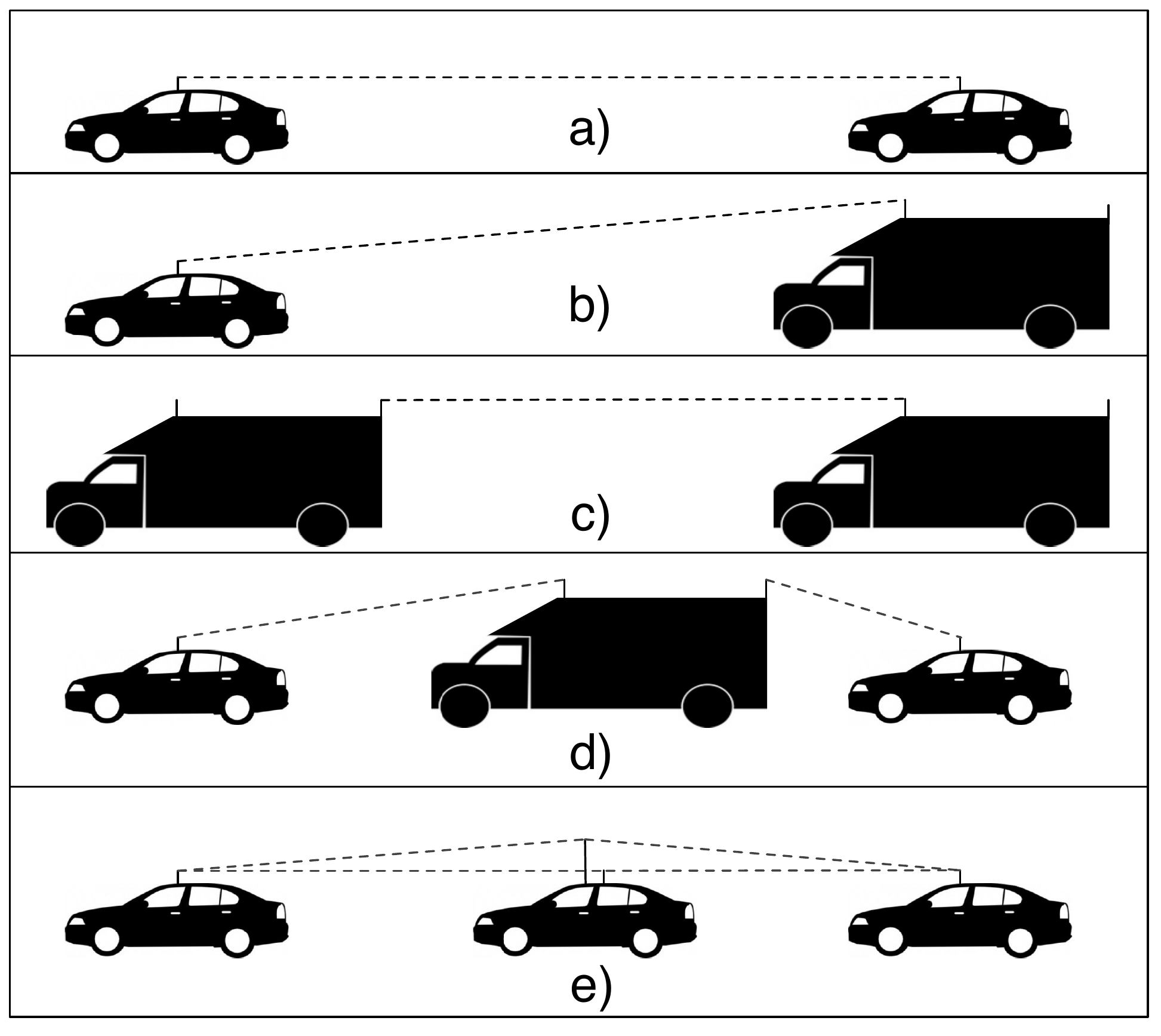}
     \caption{\small We performed the following experiments: %
      a) car-car; b) car-van; c) van-van; d) car-van-car; e) car-car-car (tall and short relay antenna).}
      \label{linksSingleMultiHop}
   \end{center}
\end{figure}

\begin{figure}
  \begin{center}
    \includegraphics[width=0.35\textwidth]{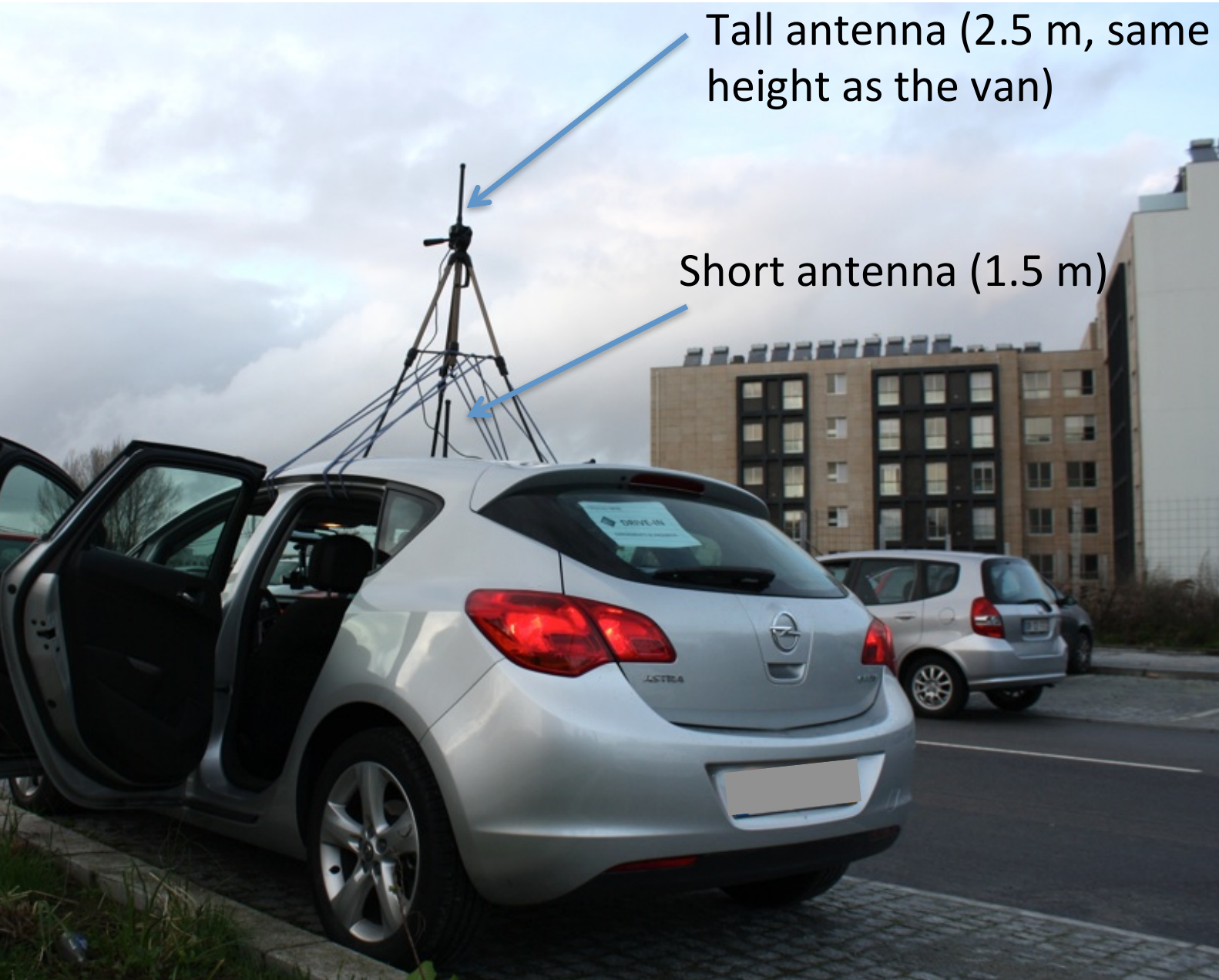}
     \caption{\small Tall and short antenna mounted on the relay vehicle. The vehicle was used as a relay node between two other short vehicles and the experiments with both antennas as relays were performed simultaneously. This experimental setup isolated the antenna height as the only factor affecting the received power and Packet Delivery Ratio (PDR). %
     We made sure that the tripod holding the tall antenna does not interfere with the short antenna by isolating any metal parts and placing the tripod legs so that they do not block the LOS with front and rear vehicle.}
      \label{twoAntennas}
   \end{center}
\end{figure}

\subsection{Experimental Scenarios}\label{subsec:scenarios}

We consider the following five scenarios. Three single-hop experiments, where two vehicles drive in tandem:
\begin{enumerate}
\item \textbf{Car-car} (Fig.~\ref{linksSingleMultiHop}a) ---  A link between two passenger cars is used to establish a baseline for single-hop comparison.

\item \textbf{Car-van} (Fig.~\ref{linksSingleMultiHop}b) --- A link between a passenger car and a van is used to evaluate the channel between vehicles of different types. %

\textbf{Van-van} (Fig.~\ref{linksSingleMultiHop}c) --- A link between two vans is used to quantify the maximum potential benefit of tall relays. When both vehicles are tall, the likelihood of their LOS being obstructed is minimized. %

\end{enumerate}
And two two-hop experiments, where three vehicles drive in tandem, the source and destination at the ends and a relay in the middle:
\begin{enumerate}\setcounter{enumi}{3}
\item  \textbf{Car-van-car} (Fig.~\ref{linksSingleMultiHop}d) --- A van is equipped with two antennas: one in front and one in back. A car drives in front of the van, exchanging messages with the van's front-mounted antenna. A second car drives behind the van, communicating with the rear-mounted antenna. This scenario quantifies the benefits of tall vehicle relays between two short vehicles.
\item  \textbf{Car-car-car} (Fig.~\ref{linksSingleMultiHop}e) --- Here we have a leading car, a trailing car and a relay car in the middle. The relay car is equipped with two radios and two antennas, one mounted directly on the roof and one mounted on a one meter tall tripod placed on top of the roof, as depicted in Fig.~\ref{twoAntennas}. This scenario enabled us to exclude the impact of all variables other than antenna height on the communication performance (i.e., the conditions in terms of terrain topography, vehicular density, and blocking vehicles were exactly the same for both tall and short antennas). %
\end{enumerate}

\begin{figure}
\centering
\subfloat[13.5~Km section of the A28 highway used in our experiments.]{\label{fig:A28}\includegraphics[height=0.35\textwidth]{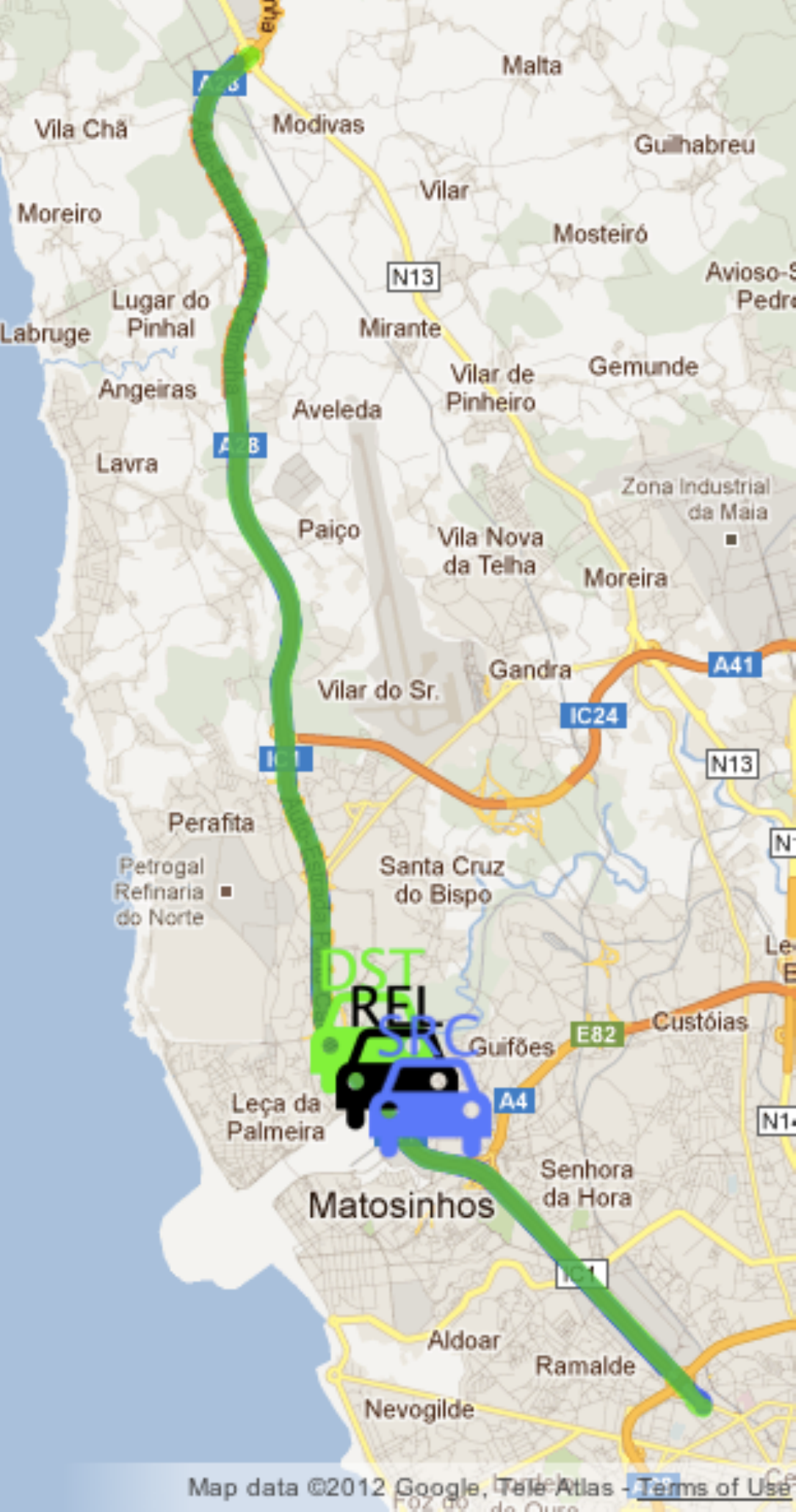}}\hspace{5mm}
\subfloat[24~Km section of the VCI urban highway used in our experiments.]{\label{fig:VCI}\includegraphics[height=0.35\textwidth]{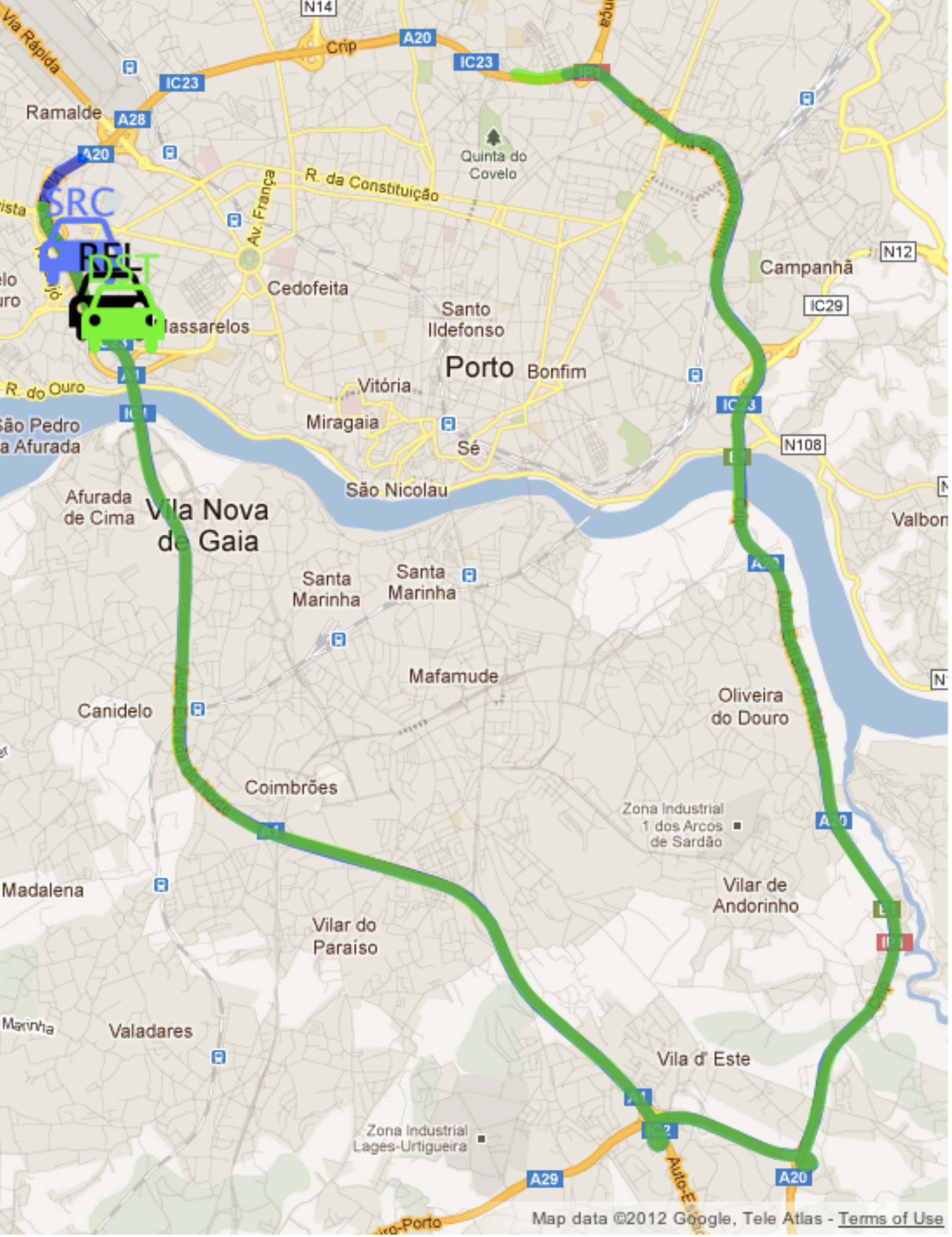}}\newline
\subfloat[Image taken from the trailing vehicle during \newline the experiments on the A28 highway.]{\label{fig:A28Img}\includegraphics[height=0.11%
\textwidth]{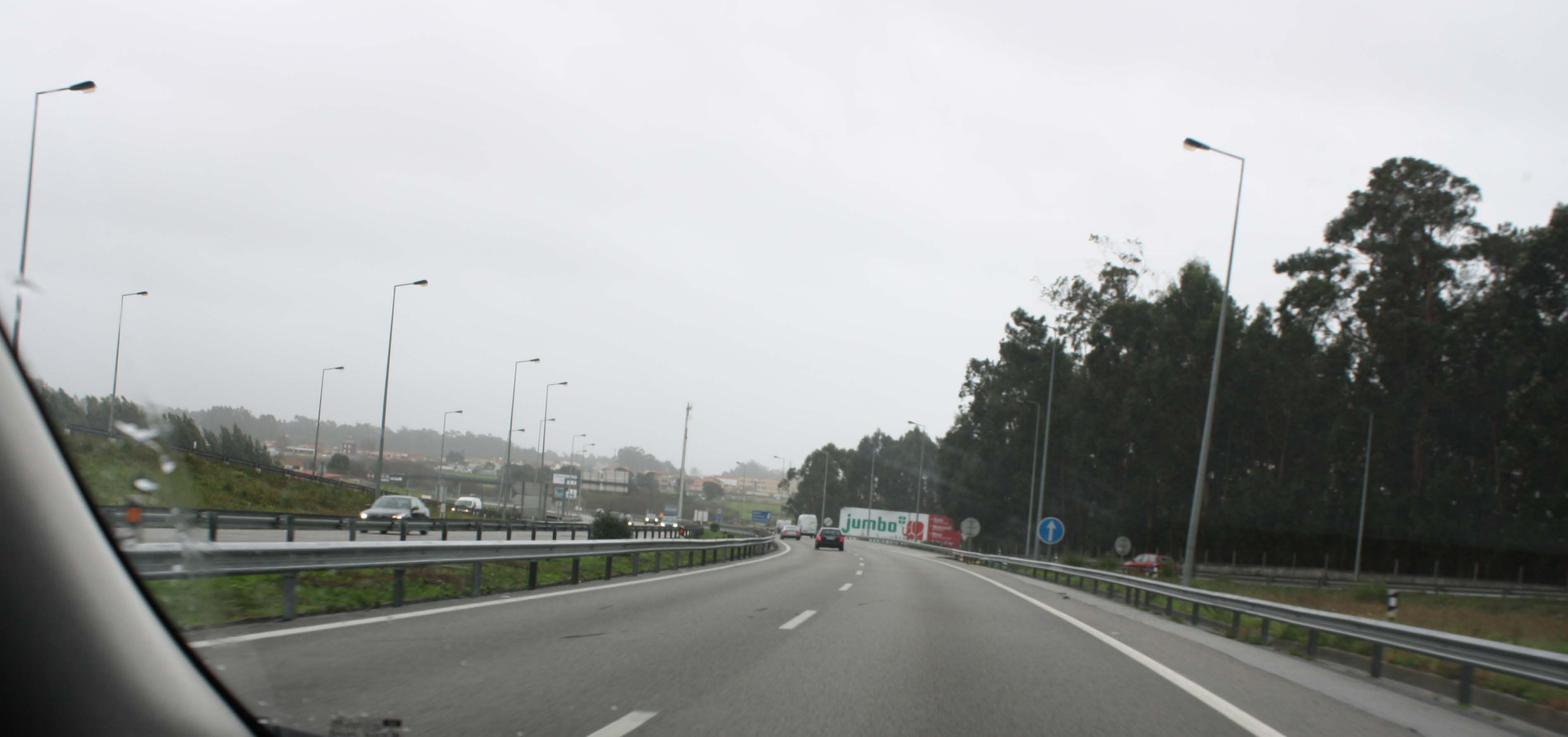}}\hspace{5mm}
\subfloat[Image taken from the trailing vehicle during the VCI highway experiments.]{\label{fig:VCIImg}\includegraphics[height=0.11\textwidth]{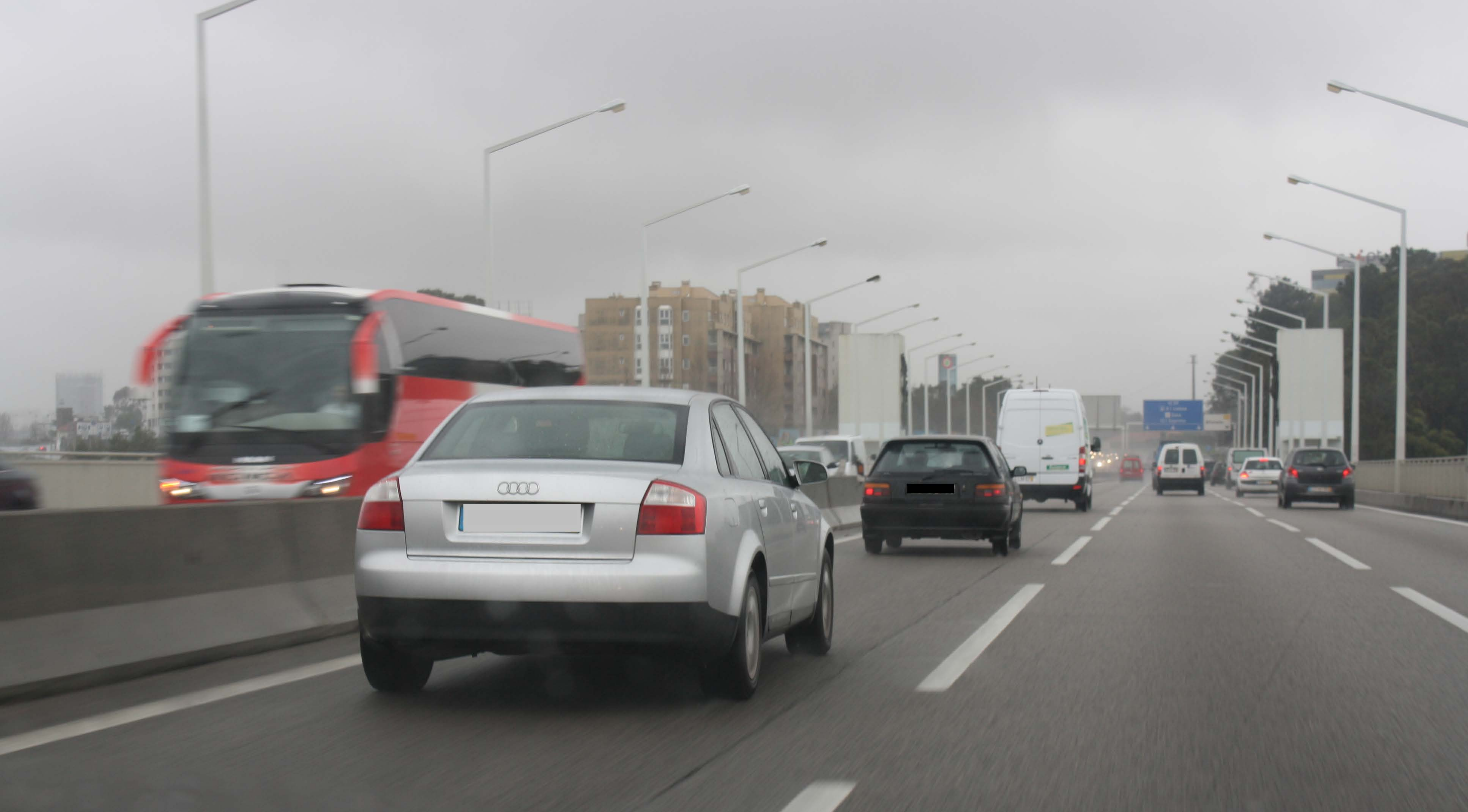}}
\caption{\small Highways where the experiments were performed. The three test vehicles shown in subfigures (a) and (b) are: source (SRC), relay (REL), and destination (DST). The SRC and DST vehicles were always passenger cars (i.e., short vehicles). The relay vehicle was either a van (tall vehicle) or a passenger car with two antennas, one mounted at 1.5~m height and the other at 2.5~m height, as shown in Fig.~\ref{twoAntennas}. The One-hop experiments were performed only on the A28 highway, whereas the two-hop experiments were performed on both highways. %
}
\label{fig:highways}
\end{figure}

Figure~\ref{fig:highways} shows the highways where we performed the experiments. The two highways, A28 and VCI, represent distinct scenarios. The A28 is a typical highway with little to no buildings near the road and occasional trees and other vegetation nearby (Fig.~\ref{fig:highways}c). The VCI highway is an urban ring road that goes around the twin cities of Porto and Vila Nova de Gaia, with occasional buildings close to the road %
and portions of the road lined with concrete walls (Fig.~\ref{fig:highways}d). 
To make the results comparable to the model-based analysis described in the previous section, we performed the experiments on the same stretch of the A28 highway that was analyzed through aerial photography (Table~\ref{dataset}). %
On both highways, the experiments were performed in %
medium to moderately dense traffic during the 11~a.m.~--~9~p.m. period on weekdays and weekends in March, April, and December, 2011. Each experiment run was approximately one hour long, with the vehicles traversing the A28 highway south to north and vice versa and making an incomplete loop on the VCI highway as shown in Fig.~\ref{fig:highways}. Speeds ranged from 40 to 120~km/h, in accordance with traffic conditions. The single-hop experiments were performed on A28, whereas the two-hop experiments were performed on both A28 and VCI.

\subsection{Hardware Setup}
Vehicles were equipped with NEC LinkBird-MXs V3, a development platform for vehicular communications~\cite{festag08}. The devices contain DSRC radios that operate in the 5.85-5.925~GHz frequency band and implement the IEEE 802.11p wireless standard~\cite{dsrc09}. Each node was configured to send periodic position messages (beacons) that were then used to record Received Signal Strength Indicator (RSSI) and Packet Delivery Ratio (PDR) information during the experiments. PDR is defined as the the ratio between the number of received messages and the number of sent messages. The reported value of RSSI is a per-packet indication of the power received over the entire 10 MHz channel during the reception of the packet's physical layer header~\cite{IEEE80211_08}. It is measured in dBm, with the radio-defined minimum noise level of -95~dBm. We validated the radios used in the experiments in an anechoic chamber: the standard deviation of the reported RSSI was under 1~dB across the radios we used. The position information was obtained from an external GPS receiver connected to each LinkBird. The system parameters are shown in Table~\ref{tab:hw-config}.

\begin{table}
	\begin{center}
		\begin{tabular}{|l c|}
			\hline
			\textbf{Parameter} & \textbf{Value} \\ 
			\hline
			\hline
			Channel Number & 180  \\ 
			\hline
			Center frequency (MHz) & 5900 \\ 
			\hline
			Bandwidth (MHz) & 10 \\ 
			\hline
			Data rate (Mbps) & 6 \\ 
			\hline
			Modulation & QPSK \\ 
			\hline
			Coding rate & 1/2 \\ 
			\hline						
			Tx power (dBm, measured) & 10 \\ 
			\hline
			Antenna gain (dBi) & 6 \\ 
			\hline
			Message frequency (Hz) & 10 \\ 
			\hline
			Message size (Byte) & 40 \\ 
			\hline
		\end{tabular}
		\caption{\small Hardware configuration parameters used for the experiments}
		\label{tab:hw-config}
	\end{center}
	
\end{table}

The radios were connected to Mobile Mark ECOM6-5500 omnidirectional antennas, which measure 26~centimeters in height. %
On the passenger cars, the antenna was positioned at the center of the roof, which has been empirically shown to be the overall optimal position~\cite{kaul07}.
On the vans, we used two antennas: one at the front of the roof, and another at the back (shown in Fig.~\ref{linksSingleMultiHop}). %
This prevents the van itself from significantly deteriorating the channel characteristics by blocking the LOS path between its own antenna and the antenna of the vehicle it is communicating with. %

\begin{figure*}
\centering

\subfloat[Experimental overall results]{\label{fig:pdr:overall}\includegraphics[width=0.28\textwidth]{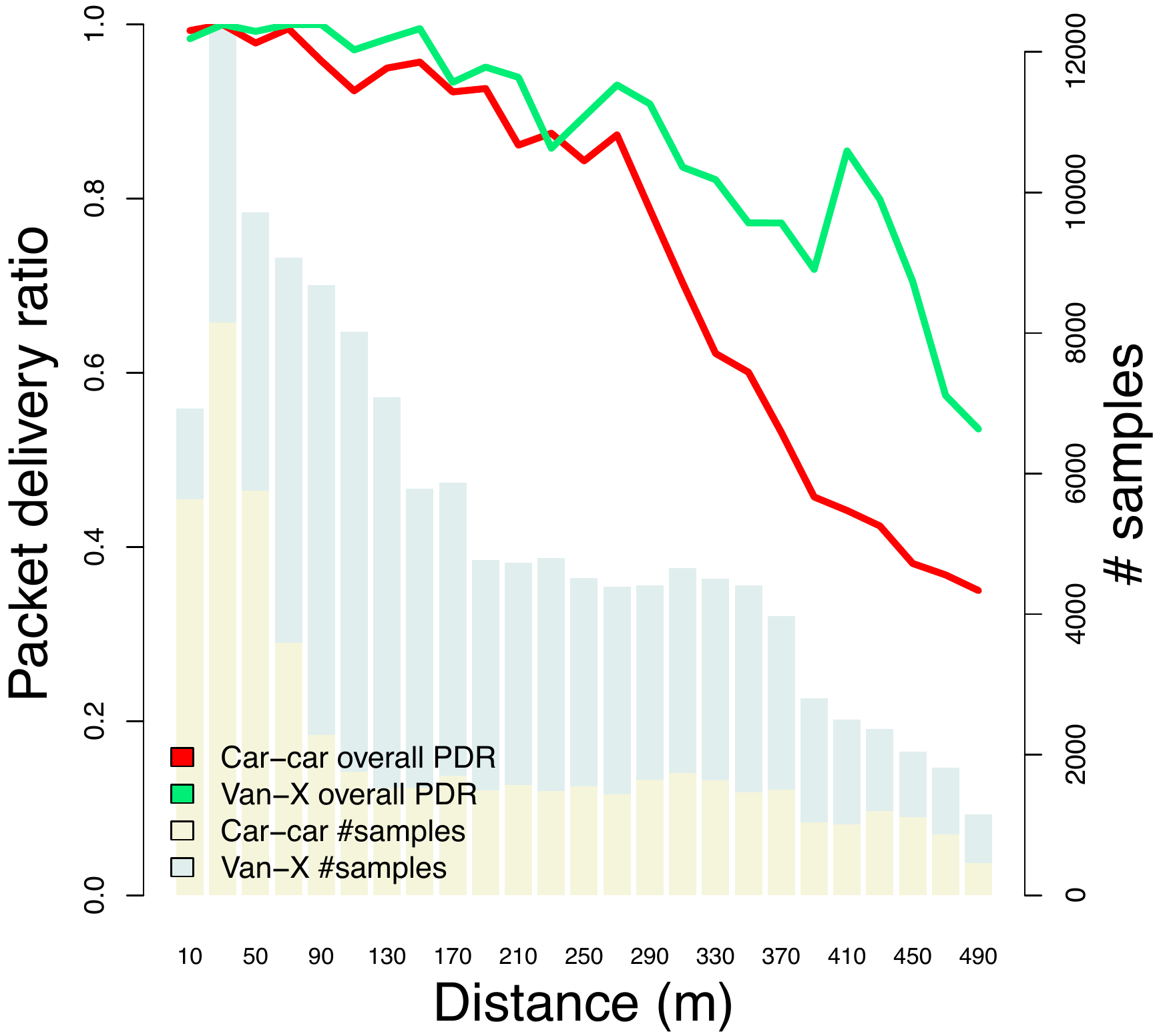}}
\subfloat[Model-based overall results]{\label{fig:pdr:overallM}\includegraphics[width=0.28\textwidth]{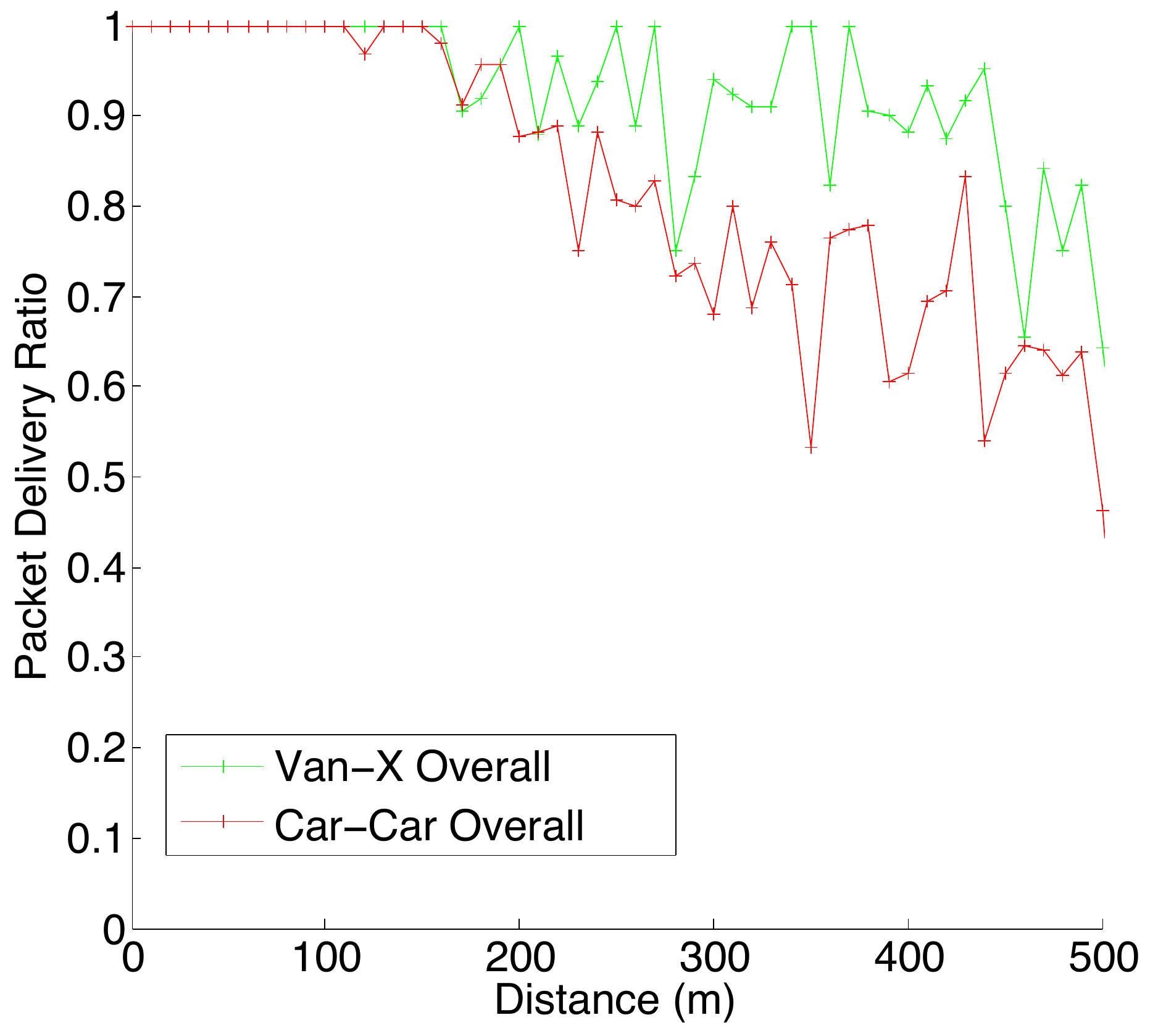}}

\subfloat[Experimental NLOS results]{\label{fig:pdr:nlos}\includegraphics[width=0.28\textwidth]{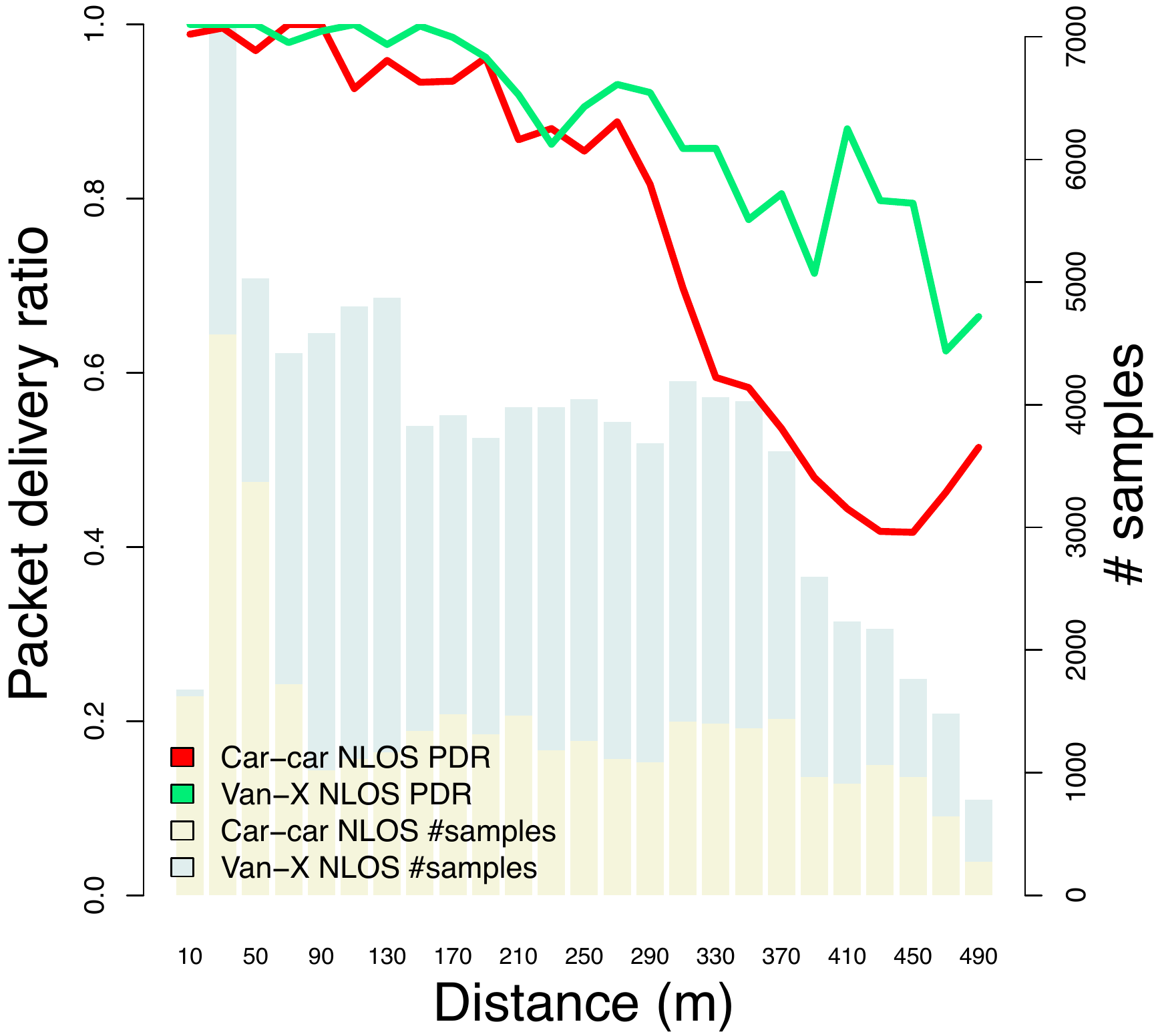}}
\subfloat[Model-based NLOS results]{\label{fig:pdr:nlosM}\includegraphics[width=0.28\textwidth]{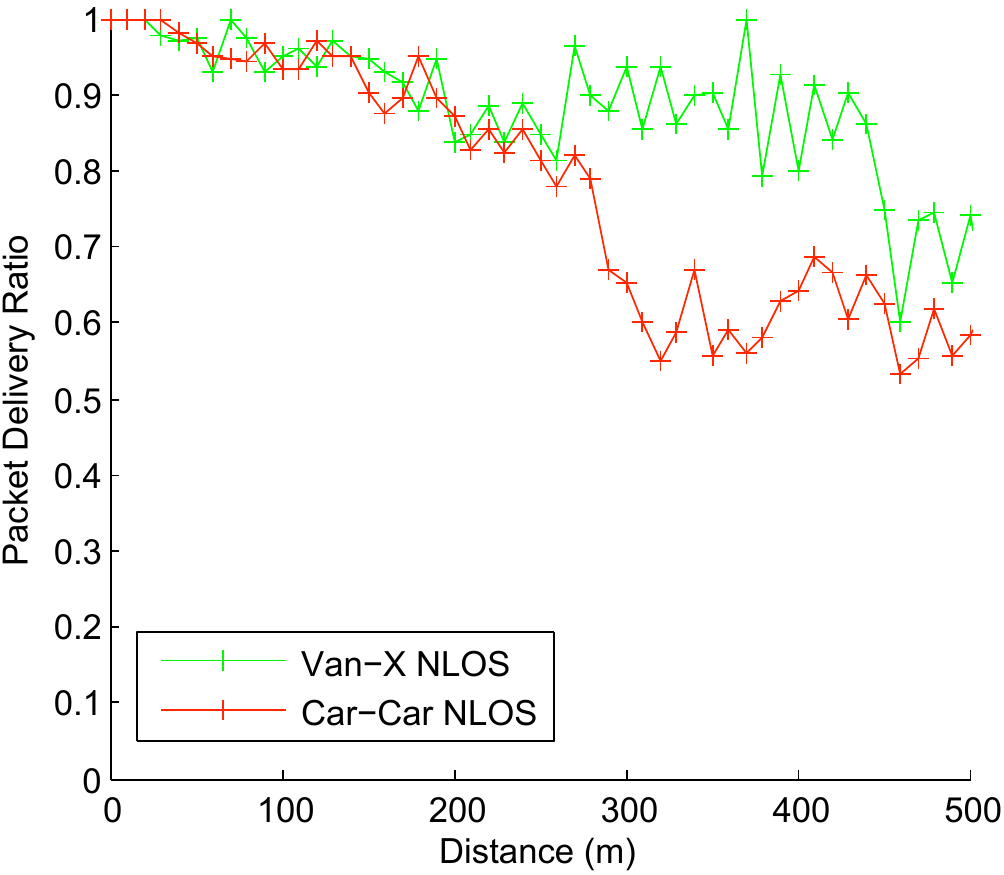}}
\caption{\small Packet Delivery Ratio (PDR) obtained through the experiments and the model for single-hop communication.} %
\label{fig:pdr}
\end{figure*}

To help us distinguish between LOS and NLOS conditions, we recorded videos of the experiments from the vehicle following in the rear in case of single-hop, and from both the leading and trailing vehicles in case of two-hop experiments (two videos were required in two-hop experiments to determine LOS conditions for each link). We then synchronized the videos to the experimental data using a custom web-based visualization suite~\cite{losexpsite} and classified each part of the experiment as LOS or NLOS with a one second resolution. We classified the conditions as NLOS when one or more vehicles, short or tall, were present between the two communicating vehicles. Given that the experiments were performed on highways, the number of static obstructions such as buildings was negligible and thus not considered.

\subsection{Experimental Results}
In this section, we present the results of the experiments we performed. We also validate the channel model developed in \cite{boban11}, which we employ in subsequent system-level simulation study of the benefits of tall vehicle relays (described in Section~\ref{sec:largeScale}).
Using the exact dimensions and locations of the vehicles, %
the model calculates additional attenuation due to vehicles. Based on the concept of multiple knife-edge attenuation described in \cite{itu07}, the model takes into account the attenuation on a radio link due to vehicles intersecting the ellipsoid corresponding to 60\% of the radius of the first Fresnel zone. Each vehicle is abstracted as a single knife edge; depending on the number and the severity of the LOS obstruction, the model calculates the additional attenuation. %

\subsubsection{One Hop Experiments}

We first present results for one-hop PDR as a function of distance, depicted in Fig.~\ref{fig:pdr}. The figure shows the PDR results obtained through both the experiments and the model described in the previous section. Similar to the model-based results, we aggregate the van-van and van-car cases to analyze the benefit of tall vehicles regardless of the height of the other vehicle. We call this combined scenario Van-X. For each message sent, we check whether it was received or not and place that information in a distance bin with a 20 meter granularity based on the distance between the communicating vehicles. In addition to the PDR, for experimental data we plot the number of samples placed in each bin.

\begin{figure*}
\centering
\subfloat[Car-Car A28 NLOS RSSI. Per-link difference
between model
and measurements:
Mean: -0.2 dB
Std. Dev: 5.6 dB]{\label{fig:1hopShortA28}\includegraphics[trim=0cm 5.5cm 0cm 5.5cm,clip=true, width=0.3\textwidth]{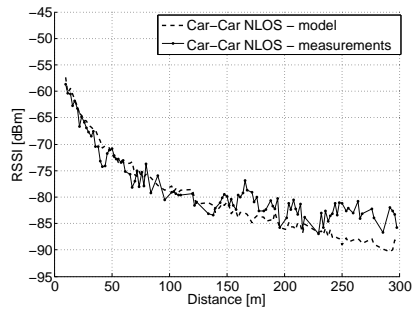}} \hspace{5mm}
\subfloat[Van-X A28 NLOS RSSI. Per-link difference
between model
and measurements:\newline
Mean: -0.2 dB
Std. Dev: 4.3 dB]{\label{fig:1hopTallA28}\includegraphics[trim=0cm 5.5cm 0cm 5.5cm,clip=true,width=0.3\textwidth]{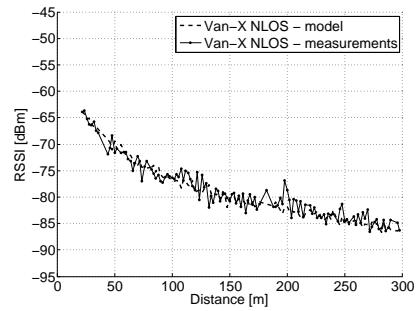}}\\
\caption{\small Received signal power obtained through the experiments and the model for single-hop communication on A28 highway. Figures show the mean  received power for two-meter distance bins. Results are plotted only for bins with at least 40 data points.}
\label{fig:RSSISingleHop}
\end{figure*}

Figure~\ref{fig:pdr:overall} shows the overall experimentally obtained PDR for both Car-Car and Van-X scenarios, regardless of the LOS conditions. We can observe that the Van-X PDR is consistently better than the Car-Car PDR. Up to 280 meters, the difference is slight but after that it becomes quite significant, with Van-X offering an improvement of around 20 percentage points over Car-Car communication up to the maximum distance for the recorded data. Figure~\ref{fig:pdr:overallM} depicts the model-derived overall PDR, based on aerial photography of the same A28 highway. The PDR exhibits a behavior similar to that of the experimentally collected data (Fig~\ref{fig:pdr:overall}). %

Figure~\ref{fig:pdr:nlos} depicts the experimentally obtained PDR for NLOS cases only, where there were other vehicles between the communicating vehicles that potentially obstructed the LOS. The shapes of the curves are similar to the overall case, %
with Van-X providing a clear advantage when compared to Car-Car communication at distances larger than 250 meters. %
When the received power is close to the reception threshold, the improved channel made possible by the use of tall vehicles often makes the difference between a decodable and a non-decodable packet. Figure~\ref{fig:pdr:nlosM} shows the PDR for NLOS data as predicted by the model. As with the overall case, the results are similar to those obtained experimentally. %
For A28 highway, Fig.~\ref{fig:RSSISingleHop} shows that the received power for NLOS links generated by the model matches well the measurements. Specifically

\begin{figure}
  \begin{center}
    \includegraphics[width=0.35\textwidth]{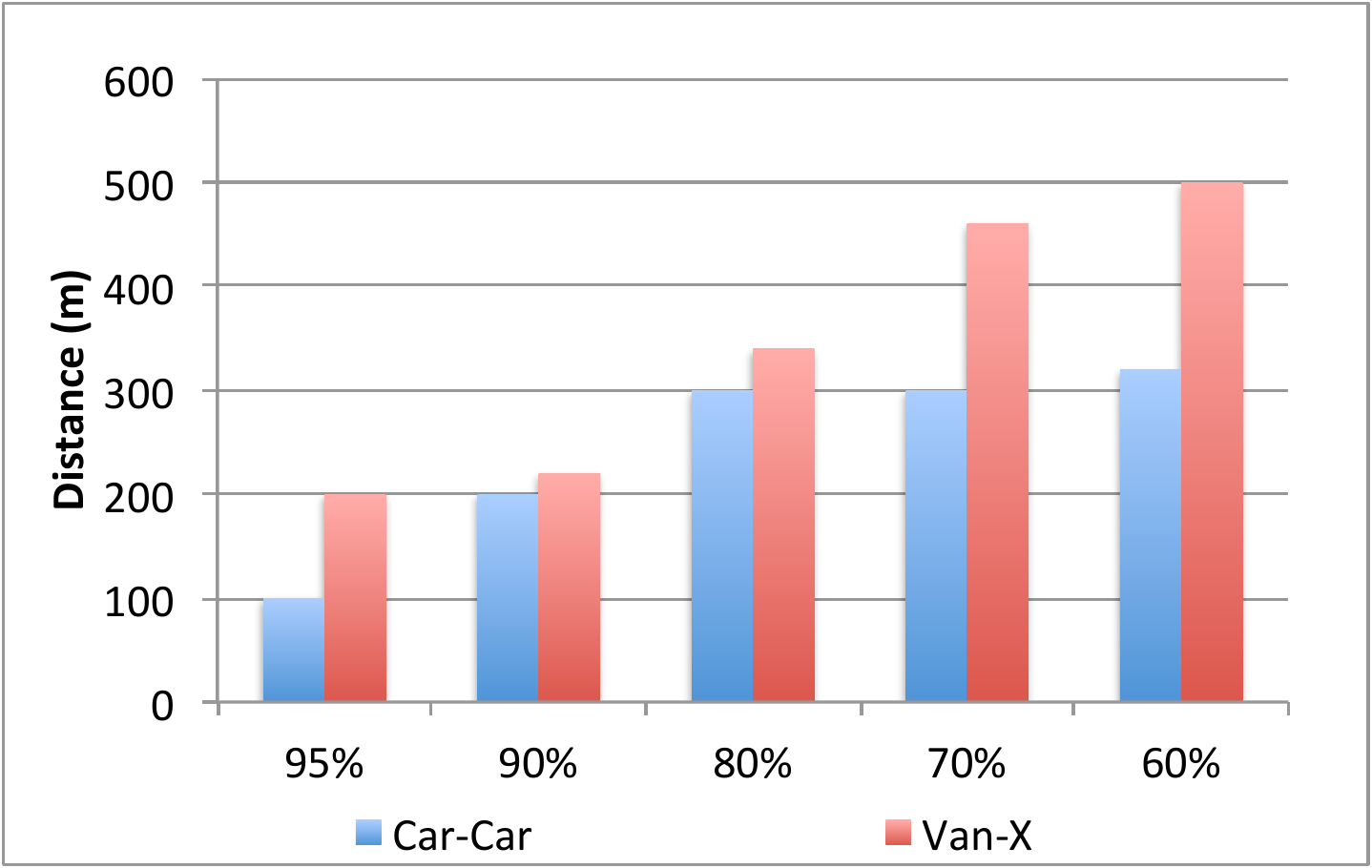}
     \caption{\small  Experimental results on the effective communication range as a function of desired packet delivery ratio for NLOS conditions.}
      \label{fig:effective-comm-range}
   \end{center}
\end{figure}

From an application's point of view, the benefit of using tall vehicles as forwarders can be seen as an increase in the effective communication range given a certain delivery probability requirement. Figure~\ref{fig:effective-comm-range} shows the difference in communication range under NLOS conditions, using the data derived from the graph in Fig.~\ref{fig:pdr:nlos}, as a function of the desired delivery ratio. Tall vehicles increased the effective communication range by a margin of up to 200 meters. %
The results show that significant benefits can be achieved by differentiating vehicles according to their height. Selecting tall vehicles allows for higher probability of LOS, increased network reachability and received signal power, all of which result in a higher PDR, which is of particular importance for effective implementation of safety applications~\cite{torrent06}.

\begin{figure*}
\centering
\subfloat[A28 Packet Delivery Ratio]{\label{fig:A28PDR}\includegraphics[width=0.3\textwidth]{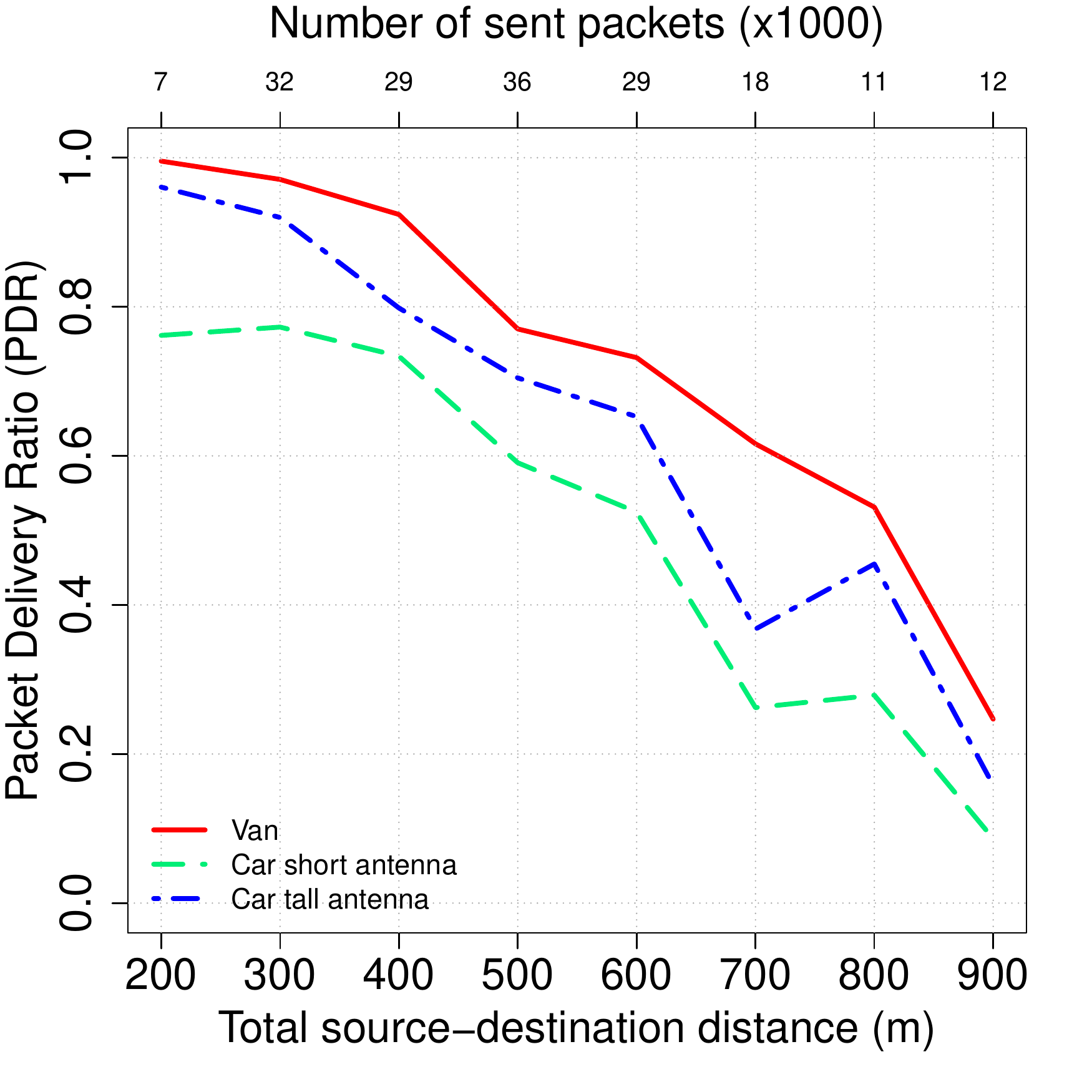}}
\subfloat[VCI Packet Delivery Ratio]{\label{fig:VCIPDR}\includegraphics[width=0.3\textwidth]{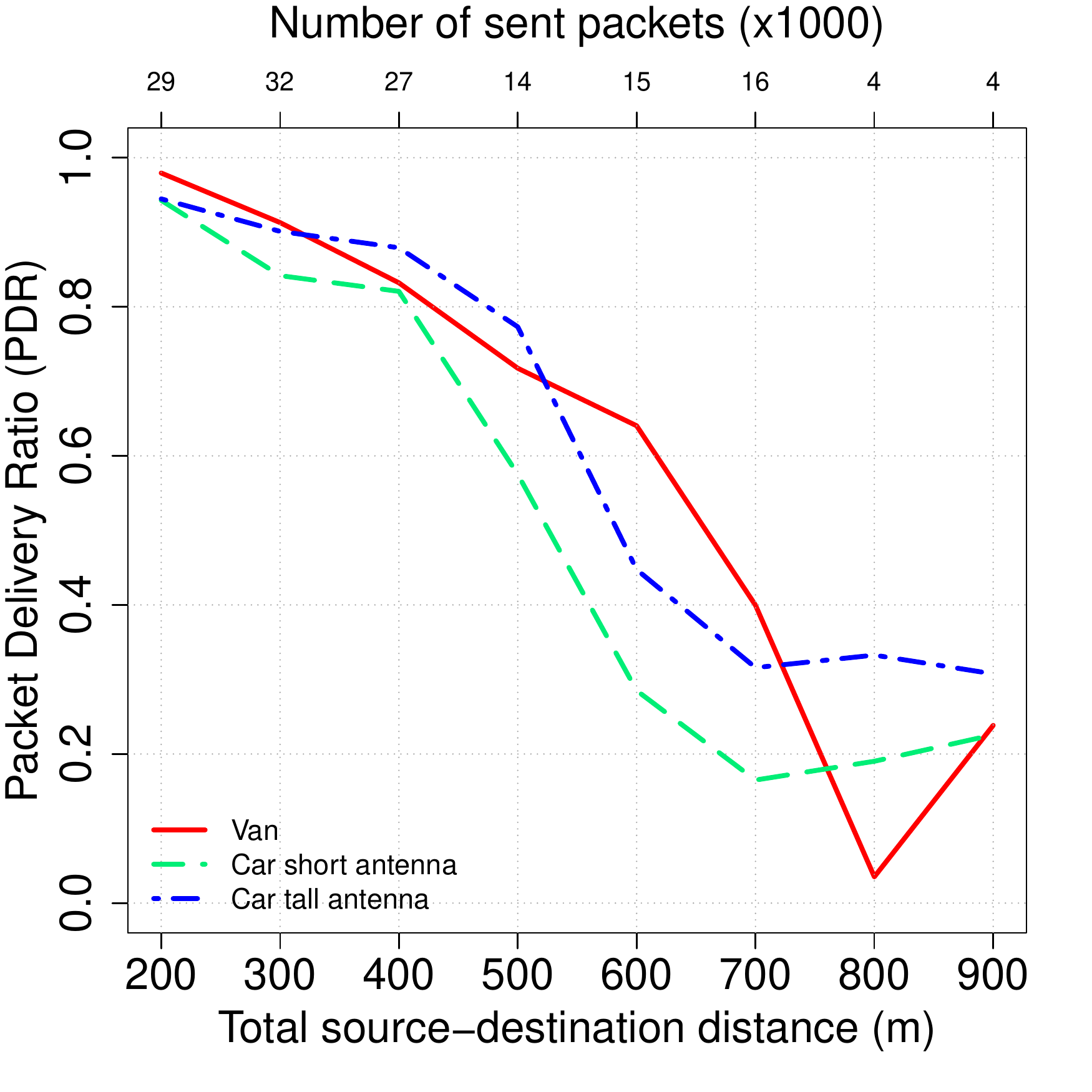}}
\caption{\small Overall Packet Delivery Ratio (PDR) results for the two-hop experiments. The end-to-end PDR is computed by multiplying the PDR of the two individual links.}
\label{fig:2hopPDR}
\end{figure*}

\begin{figure*}
\centering
\subfloat[A28 Overall (aggregated LOS and NLOS) RSSI]{\label{fig:2hop-rssi-ExpVsModA28}\includegraphics[width=0.3\textwidth]{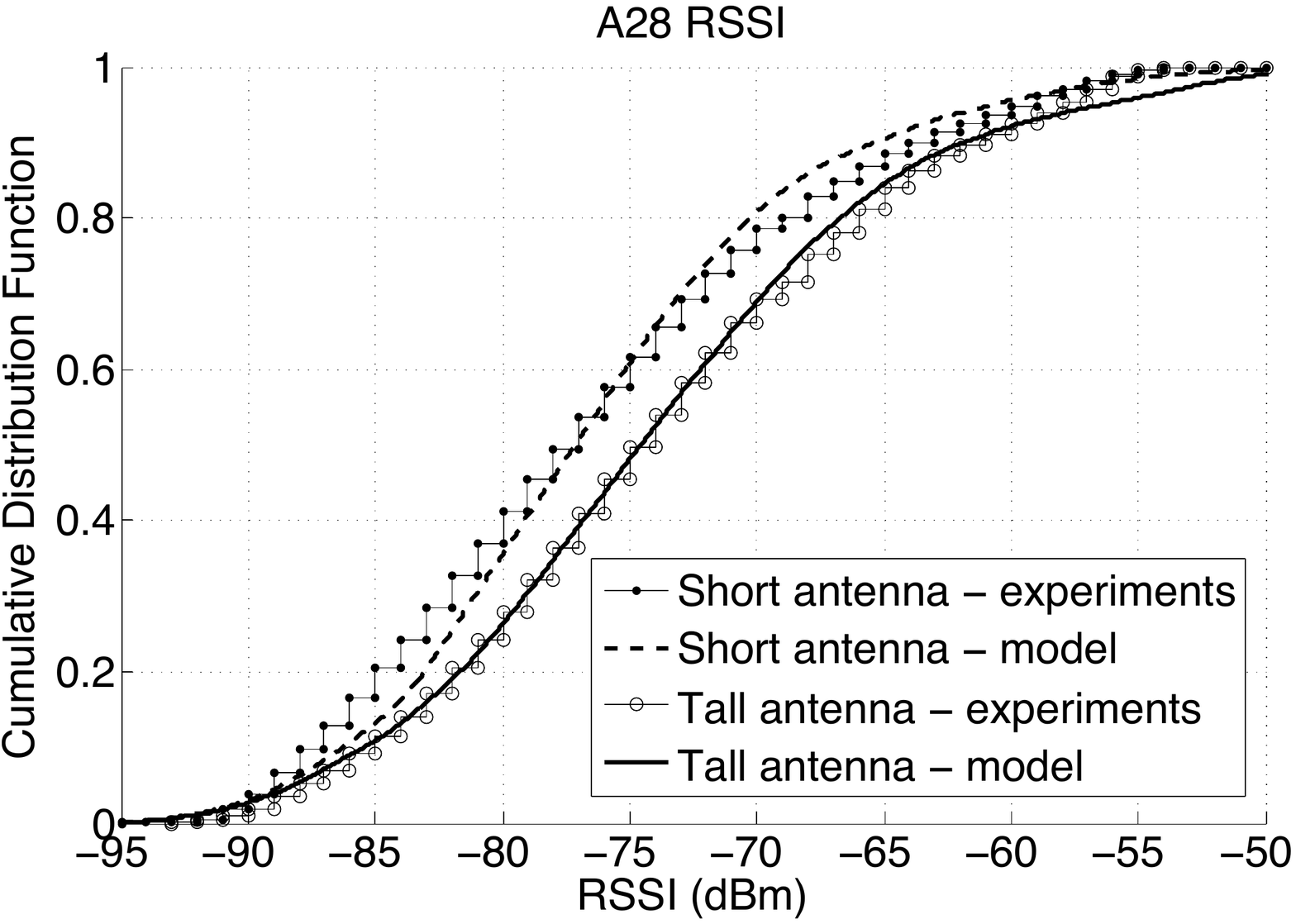}} \hspace{5mm}
\subfloat[VCI Overall (aggregated LOS and NLOS) RSSI]{\label{fig:2hop-rssi-ExpVsModVCI}\includegraphics[width=0.3\textwidth]{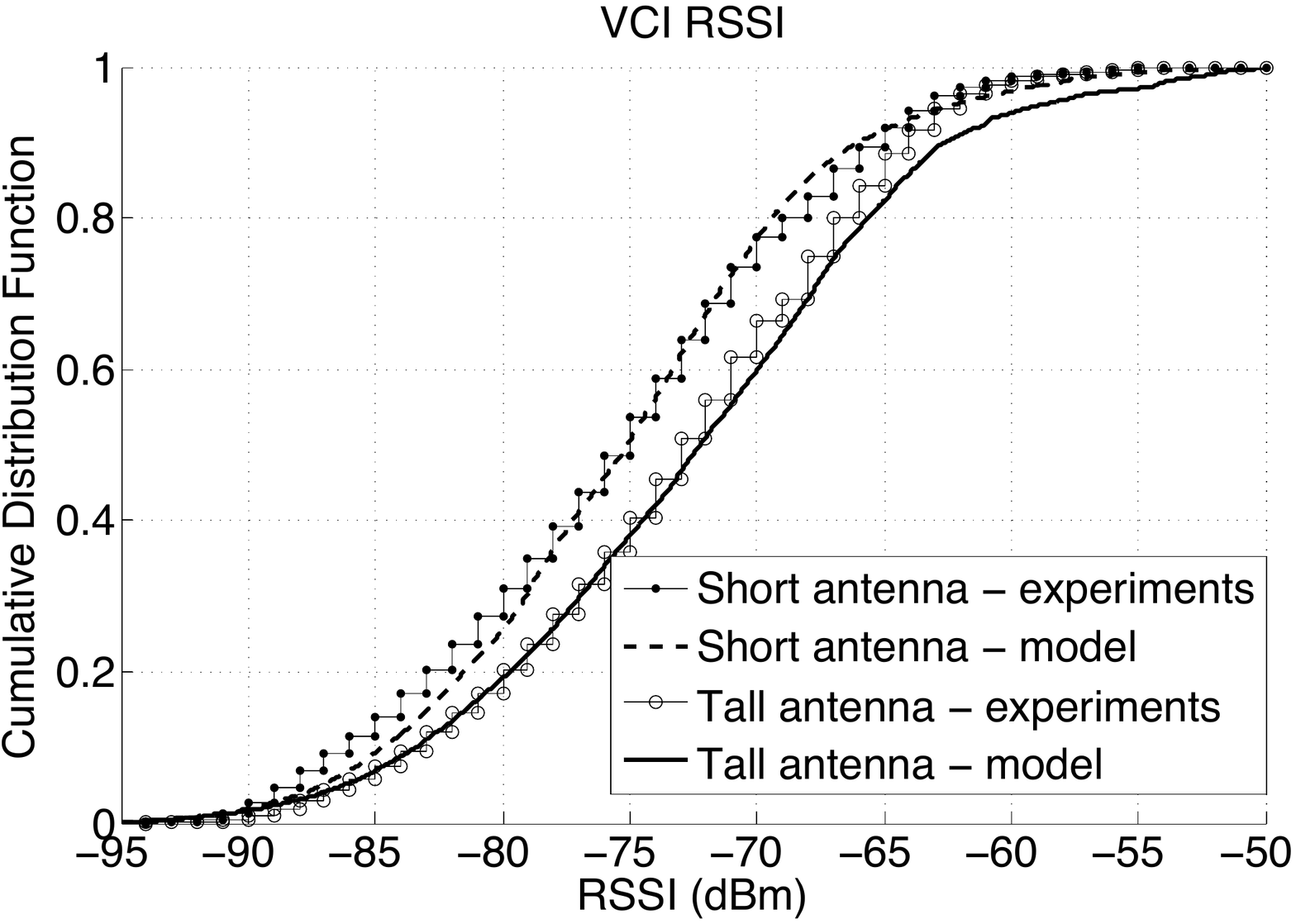}}\\
\caption{\small Cumulative Distribution Functions of the Received Signal Strength Indicator (RSSI) for the tall and short relay antennas, for both the car-car-car experiments (Fig.~\ref{linksSingleMultiHop}e) and the channel model. Both the LOS data (i.e., no obstruction) and NLOS data (i.e., vehicle obstructions) is included. LOS data comprises 66\% of the total data, with the remaining 34\% is NLOS due to vehicles.}
\label{fig:RSSICDF}
\end{figure*}

\begin{figure*}
\centering
\subfloat[NLOS A28 RSSI gains]{\label{fig:2hop-rssi-gains-a28}\includegraphics[width=0.3\textwidth]{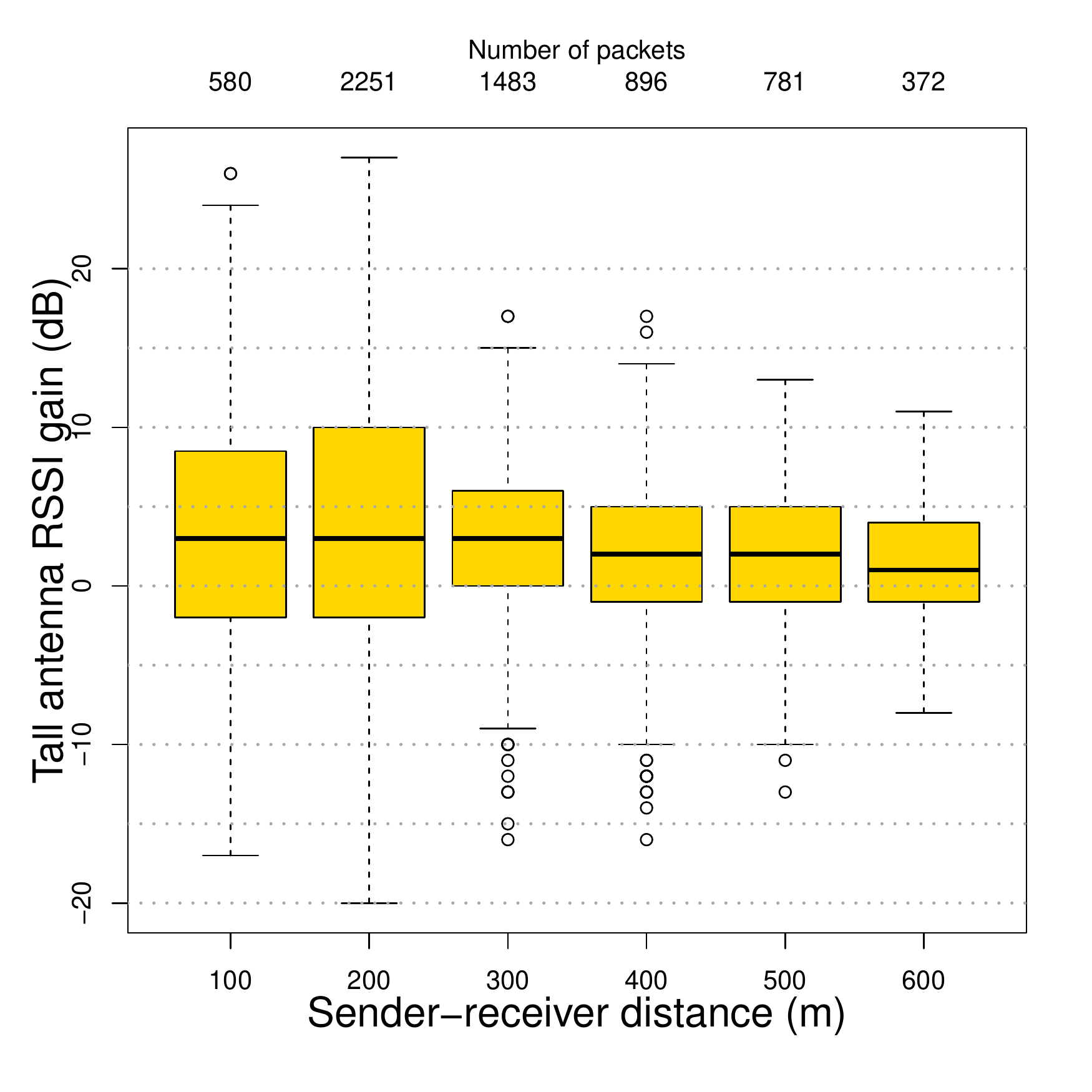}}
\subfloat[NLOS VCI RSSI gains]{\label{fig:2hop-rssi-gains-vci}\includegraphics[width=0.3\textwidth]{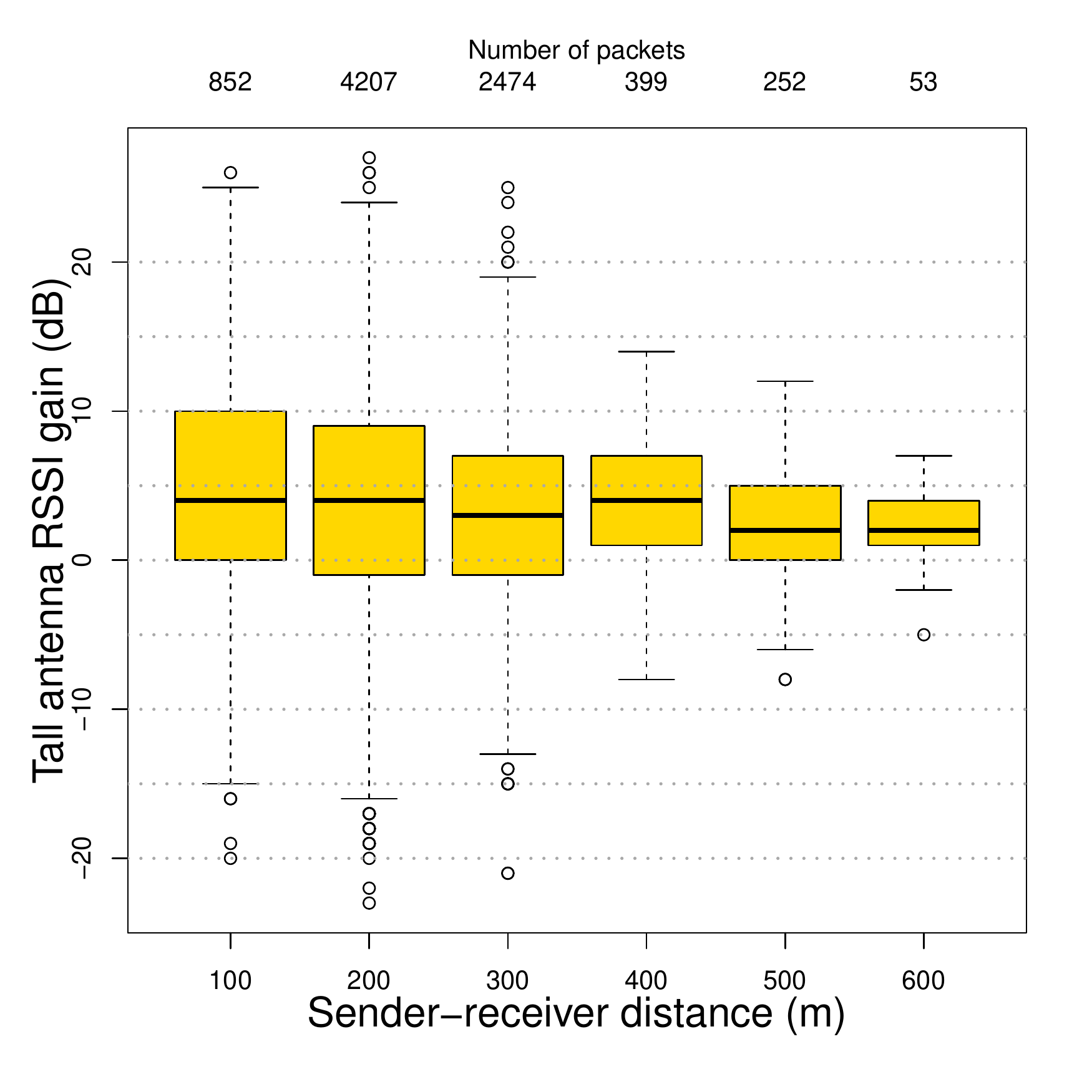}}
\caption{\small %
RSSI gains from the tall relay antenna relative to the short relay antenna for the car-car-car experiments (Fig.~\ref{linksSingleMultiHop}e) under non-LOS (NLOS) conditions. Each box plot represents the median and lower and upper quartiles. The error bars represent the minimum and maximum ranges, except for outliers (more than 1.5 times the interquartile range), which are represented by small circles.}
\label{fig:2hop-rssi-gains}
\end{figure*}

\subsubsection{Two Hop Experiments}

Figure~\ref{fig:2hopPDR} shows the overall (i.e. aggregated LOS and NLOS) end-to-end
PDR results obtained for the two-hop experiments on the A28 and VCI highways (Figs.~\ref{linksSingleMultiHop}d and \ref{linksSingleMultiHop}e). There are three PDR curves for each highway: 1) for the car-van-car scenario (Fig.~\ref{linksSingleMultiHop}d); 2) for the car-car-car scenario using the low-mounted antenna as a relay (Fig.~\ref{linksSingleMultiHop}e), and 3) for the car-car-car scenario using the high-mounted antenna as the relay (Fig.~\ref{linksSingleMultiHop}e). Curves 2) and 3) share the exact same spatial and temporal conditions (vehicle density, surroundings, obstructing vehicles), whereas curve 1) was obtained by redoing the experiments with a van as a relay. 

The PDR results follow a trend similar to the one-hop results (Fig.~\ref{fig:pdr}), with both the van and the high-mounted antenna outperforming the low-mounted antenna as relays. The taller antenna results in an improvement of up to 20 percentage points when compared with the short antenna. Using a van results in an even more pronounced improvement of up to 40 percentage points at larger communication distances. Also, note that these results indirectly confirm the potential of using tall relays decreasing for end-to-end delay: higher PDRs will reduce the number of retransmissions required, which will save time.

Fig.~\ref{fig:RSSICDF} shows the RSSI Cumulative Distribution Function (CDF) for the car-car-car two-hop experiment (Fig.~\ref{linksSingleMultiHop}e), where the relay vehicle has both tall and short antennas, and the RSSI values generated by the channel model using the vehicle location information and LOS conditions obtained during the experiments. The plots encompass the aggregated data for LOS and NLOS due to vehicles. %
The tall relay antenna shows a consistent advantage over the short antenna, with up to 4~dB higher RSSI. Furthermore, there is a good agreement between the experimental and model-derived values.

To obtain a deeper insight into the benefits of a tall antenna in NLOS conditions,  
Fig.~\ref{fig:2hop-rssi-gains} %
 shows the Received Signal Strength Indicator (RSSI) %
results in the form of a box plot for each 100~meter sender-receiver distance bin in the case of NLOS communication due to vehicles. %
We computed difference in received power for the pairs of packets that were received by both the high and the low-mounted antennas. %
The high-mounted antennas provide a median gain between 2 and 4~dB in received power. As discussed earlier, the benefit is due to the higher-mounted antenna being less susceptible to LOS blocking from the non-communicating vehicles.

\section{TVR -- Tall Vehicle Relaying Technique}
\label{sec:TVR}

Having analyzed the potential benefits of tall vehicles as relays, we present a heuristic that allows routing schemes to capitalize on this opportunity. We focus on geographic routing schemes -- contrary to traditional topology-based solutions, in geographic routing paths are constructed on-the-fly, based on the geographical location of the nodes. This makes it especially suitable for vehicular environments, which are characterized by highly dynamic topologies. In geographical routing, at each hop, the node holding the packet will choose one node from its neighborhood to act as a relay for the packet. Our goal is to provide an heuristic that leads to good next hop relay selection. In the subsequent text, we make the following assumptions:
\begin{itemize}
\item The destination of a packet is specified by a set of geographical coordinates.
\item Vehicles make use of a location system such as GPS.
\item \emph{Neighbor} is defined as a vehicle which receives the signal from the current vehicle above the sensitivity threshold, based on the employed channel model. %
\item Vehicles transmit periodic beacons with their location; this allows each node to build a neighbor table with the location of its neighbors. The beacons also include a binary variable that says whether the transmitting node is tall or short.
\end{itemize}

Tall vehicle experiments shown in Figs.~\ref{fig:pdr}~and~\ref{fig:effective-comm-range} show that, on average, tall vehicles have a larger communication range. Therefore, when choosing a relay to minimize the number of hops, a tall vehicle is preferable to a short vehicle if they are equidistant from the current transmitter. The same is true if the tall vehicle is closer to the destination than the short vehicle. If, on the other hand, the short vehicle is closer to the destination, then the tall vehicle is only preferable if the range improvement that it provides is enough to offset the initial distance advantage of the short vehicle. With this in mind, we propose the TVR heuristic, which works as follows:

\begin{enumerate}

\item The neighbors in the direction of the destination are divided into tall and short neighbors, according to their heights.

\item The farthest neighbor from each subset, $Far_{Short}$ and $Far_{Tall}$, are computed according to: %

\begin{equation}
\arg\min_{x\in \mathcal{N} (Tx)} dist(x, d),
\label{eq:farthestNeighbor}
\end{equation}
 where $dist(x,d)$ is the Euclidean distance between neighbor $x$ and the destination $d$, $Tx$ is the current transmitter %
and $\mathcal{N}(x)$ is the set of neighbors of $x$. %

\item If $dist(Tx,Far_{Short})-dist(Tx,Far_{Tall})\leq x_{max}$, $Far_{Tall}$ is selected; otherwise $Far_{Short}$ is selected.

\end{enumerate}

In other words, TVR selects a tall vehicle if the distance difference between the farthest tall vehicle and the current transmitter and the farthest short vehicle and the transmitter is less or equal than a threshold $x_{max}$; otherwise, the farthest short node is selected. 

\begin{figure*}
  \begin{center}
    \includegraphics[width=0.88\textwidth]{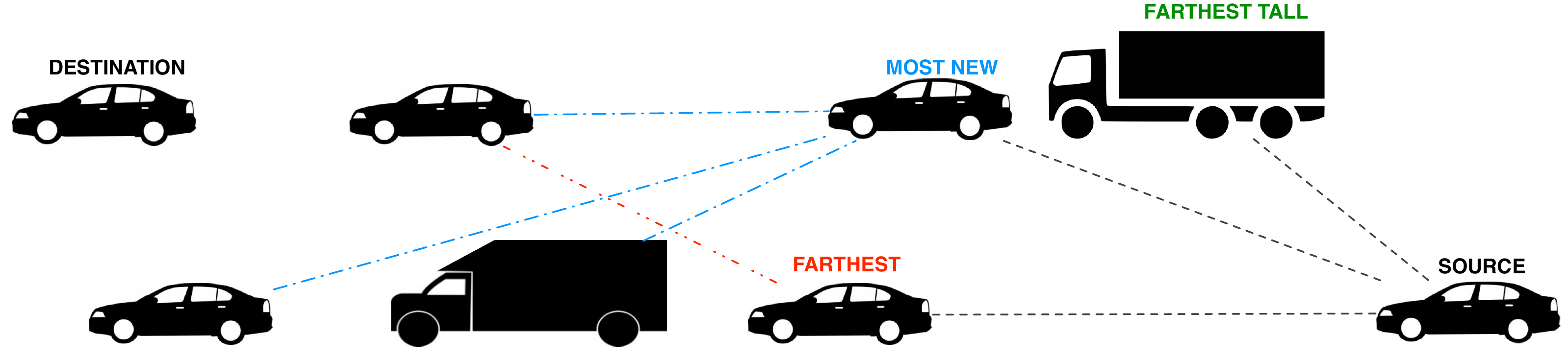}
     \caption{\small Relay selection for the three techniques. In case of \emph{Most New Neighbors} relay technique, the vehicle designated \emph{Most New Neighbors} will be selected, as it has most new neighbors (three) in the direction of the destination (designated \emph{Destination}) that are not neighbors of the current sending node (designated \emph{Source}). In case of \emph{Farthest Neighbor} relay technique, vehicle designated \emph{Farthest Neighbor} will be selected, as it is farthest from the current sending node, and in the direction of the destination \emph{Dest}. In the case of TVR, the tall vehicle designated \emph{Farthest Tall} will be selected. Note that a single vehicle can be selected by multiple techniques (e.g., farthest vehicle might have most new neighbors, and it can also be a tall vehicle, which would make it the best relay for all three techniques)}
      \label{MostNew-Tall-Farthest}
   \end{center}
\end{figure*}

\subsection{Calculating $x_{max}$}
\begin{figure}
  \begin{center}
    \includegraphics[width=0.45\textwidth]{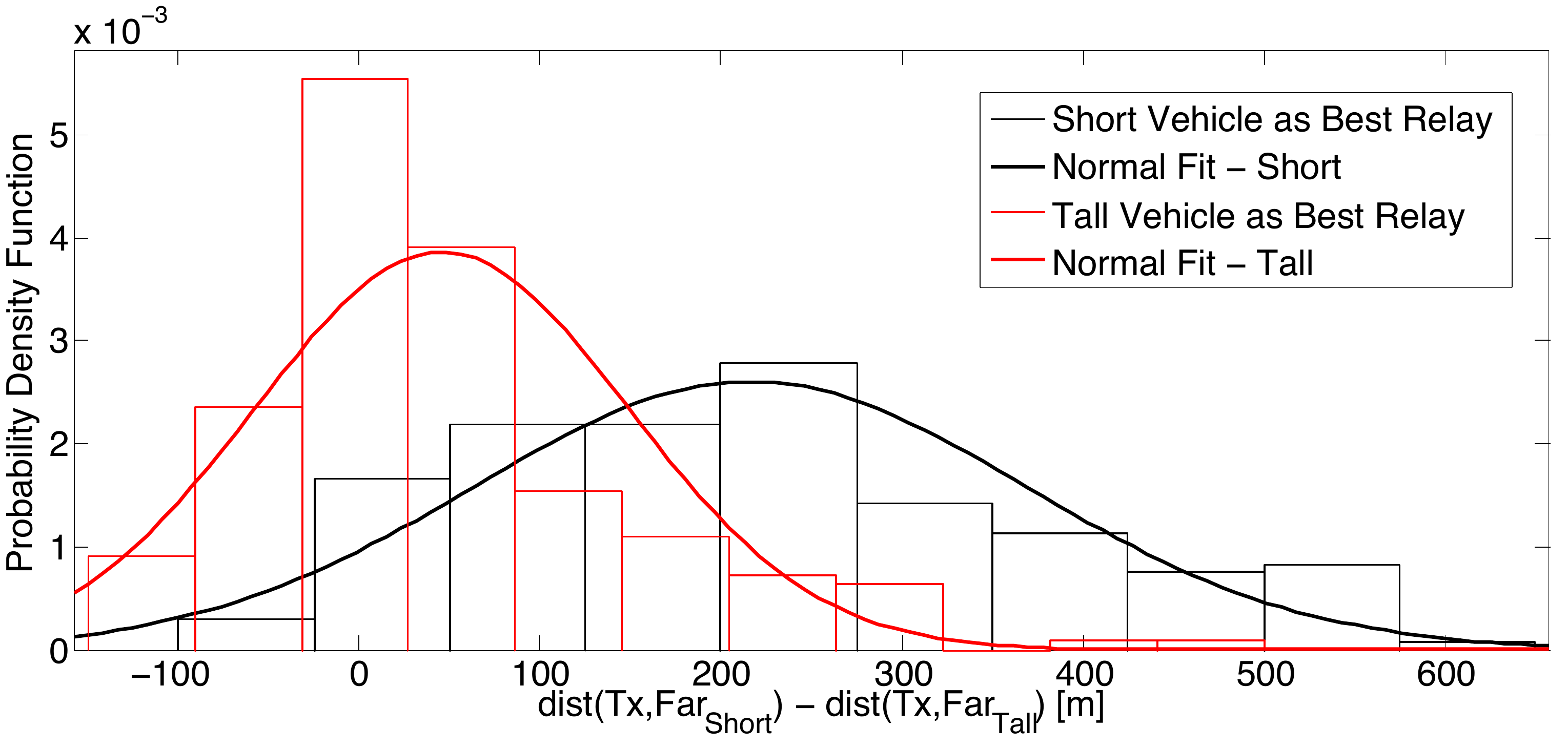}
     \caption{\small Probability distributions of the distance from the transmitter to the farthest short and farthest tall vehicle $dist(Tx,Far_{Short})-dist(Tx,Far_{Tall})$ for a transmit power of 10~dBm, tested on the aerial photography data of the A28 highway. %
 Negative distances implies that the tall vehicle is farther from the transmitter than the short vehicle. For the given transmit power, when a short vehicle is the best relay, it is on average 210~meters farther from the transmitter than the tall vehicle. When a tall vehicle is the best relay, it is on average 50~meters closer to the transmitter than the short vehicle.} %
      \label{tallVsShort}
   \end{center}
\end{figure}

In order to calculate  $x_{max}$, we first look at the distribution of distance difference $dist(Tx,Far_{Short})-dist(Tx,Far_{Tall})$, as shown in Fig.~\ref{tallVsShort}, which was derived from aerial photography data (Table~\ref{dataset}).
The case when a tall vehicle is the best relay (in terms of least number of end-to-end hops) is the distribution colored red, whereas the case when a short vehicle is the best relay is colored black. To determine when a tall vehicle is more likely to be a better relay, let us define a binary random variable $\theta$ as being one when a tall vehicle is more likely to be a better relay, and zero otherwise: %

\begin{equation}
\theta = \left\{ \begin{array}{cl}1, & \textrm{when } \displaystyle \frac{\int_{-\infty}^{\textstyle x} f_T(t)dt}{\int_{\textstyle x}^{+\infty} f_S(s)ds}>1;\\
0, & \textrm{otherwise,}
\end{array} \right.
\label{eq:Ptheta}
\end{equation}
where $f_T(t)$ and $f_S(s)$ are probability distributions  of $dist(Tx,Far_{Short})-dist(Tx,Far_{Tall})$ %
for best tall vehicle and best short vehicle case, respectively. In other words, we can interpret eq.~\ref{eq:Ptheta} as $\theta=1$ when the cumulative distribution $F_T(t)$ for a given value $x$ is larger than the complementary cumulative distribution of $F_S(s)$ and $\theta=0$ otherwise.

In order to calculate the maximum distance difference $x_{max}$ at which a tall vehicle is still a better relay, we need to solve $F_T(t) = 1- F_S(s)$.
In the specific case of our collected data, %
for tractability purposes we approximate the distance difference distributions of $s$ and $t$ with normal distributions (normal fits shown in Fig.~\ref{tallVsShort}). In this case, $x_{max}$ can be calculated by solving

\begin{equation}
1 - Q\left(\frac{x_{max}-\mu_s}{\sigma_s}\right) = Q\left(\frac{x_{max}-\mu_t}{\sigma_t}\right),%
\label{eq:Qfn}
\end{equation}
where $\mu_s$, $\sigma_s$, $\mu_t$, and $\sigma_t$ are the means and variances of $s$ and $t$, respectively, and $Q(\cdot)$ is the $Q$-function, defined as $Q(x) = \frac{1}{\sqrt{2\pi}} \int_x^\infty \exp\Bigl(-\frac{u^2}{2}\Bigr) \, du$. 

Figure~\ref{tallVsShort} shows the distributions of $s$ and $t$ for a single transmit power (10~dBm); to analyze the behavior of $s$ and $t$ with different communication ranges, we vary the transmission power from 1 to 20~dBm.
Distributions of $s$ and $t$ are readily available in simulators by implementing an appropriate channel model (such as~\cite{boban11}), since the global network knowledge (``oracle'') is available. However, 
obtaining these distributions  %
is not straightforward without global knowledge, %
which means that %
the distributions of $s$ and $t$ will not be available to the routing protocols in vehicles. Therefore, we %
set a fixed value for $x_{max}$. %
 We used a value of $x_{max}$ calculated based on the aerial photography dataset in Table~\ref{dataset} as follows. We choose $x_{max}$ to be the average value of $t$ across %
 transmission powers from 1 to 20~dBm (typical transmit powers for the DSRC standard). %
Formally, 
\begin{align}   \label{eq:proradi}
x_{max}&= \sum_{i=1}^{20}E[t|Pwr=i~dBm]\cdot P[Pwr=i~dBm] \\%E[t_{overall}]\\
\nonumber&=\frac{1}{20}\sum_{i=1}^{20}E[t|Pwr=i~dBm]\\
\nonumber&=\frac{1}{20}\sum_{i=1}^{20}\int_{-\infty}^{\infty}tf_T(t|Pwr=i~dBm)dt,
\end{align}   
where $Pwr$ is the transmit power.
The calculated value is $x_{max} = 50$~meters (i.e., in the simulations, we use a tall vehicle %
as the next hop when $dist(Tx,Far_{Short})-dist(Tx,Far_{Tall})\leq 50$). Note that calculating $x_{max}$ based on specific values of $E[t|Pwr]$ yields better results for that specific transmission power. However, using different values of $x_{max}$ might be impractical for protocol implementation, as it may vary across different environments. %

\subsection{Other Relay Techniques Under Consideration}\label{subsec:techniques}
Ideally, we would compare the performance of TVR with the optimal relay selection technique, one that analyzes all possible end-to-end routes and selects the best composite route. However, the difficulty with this approach is that, for the number of scenarios we analyzed, employing the optimal relaying scheme was computationally infeasible. Therefore, we evaluated the performance of TVR against two existing relay techniques that make relaying decisions based on local information.  
In the subsequent text, we define a \emph{neighbor} as a vehicle which receives the signal from the current vehicle above the sensitivity threshold, based on the employed channel model~\cite{boban11}. %

\subsubsection{Most New Neighbors \emph{technique}} 
This technique will select the neighbor that contributes most new neighbors in the direction of the destination, which are not neighbors of the current sending node. More precisely, the chosen neighbor satisfies: 

\begin{equation}
\arg\max_{x\in \mathcal{N}_d (Tx)} \left| \mathcal{N}_d(x) \setminus \mathcal{N}_d(Tx)  \right|,
\label{eq:mostNewNeighbors}
\end{equation}
where $Tx$ is the current transmitter %
and $\mathcal{N}_d(x)$ is the set of neighbors of $x$ that are closer to the destination than $x$ itself.
Note that %
this technique requires nodes to include their neighbor set $\mathcal{N}_d$ in the period beacons.

The reasoning behind this technique is that the neighbor with most new neighbors has the highest local connectivity (or, in other words, highest degree distribution) in the direction of the destination. The conjecture is that more potential next hop relays translate into a higher probability of delivery to the destination. Referring to Fig.~\ref{MostNew-Tall-Farthest}, the selected vehicle (\emph{Most New}) has most new neighbors (three) that are not neighbors of the current transmitter (\emph{Source}).

\subsubsection{Farthest Neighbor \emph{technique}} 
This technique simply selects the farthest neighbor in the direction of the destination. More precisely, the current transmitter $Tx$ selects the neighbor $x$ that satisfies Eq.~\ref{eq:farthestNeighbor}.
Referring to Fig.~\ref{MostNew-Tall-Farthest}, the selected vehicle is designated as \emph{Farthest}. The intuition behind this heuristic is that maximizing the distance travelled in each hop will lead to a smaller number of hops to reach the destination. This technique has often been used in the literature (e.g., see~\cite{wisit07}). %

\section{Evaluating the System-level Benefits of Tall Vehicle Relaying}\label{sec:largeScale} %
\begin{figure}
  \begin{center}
    \includegraphics[width=0.3\textwidth]{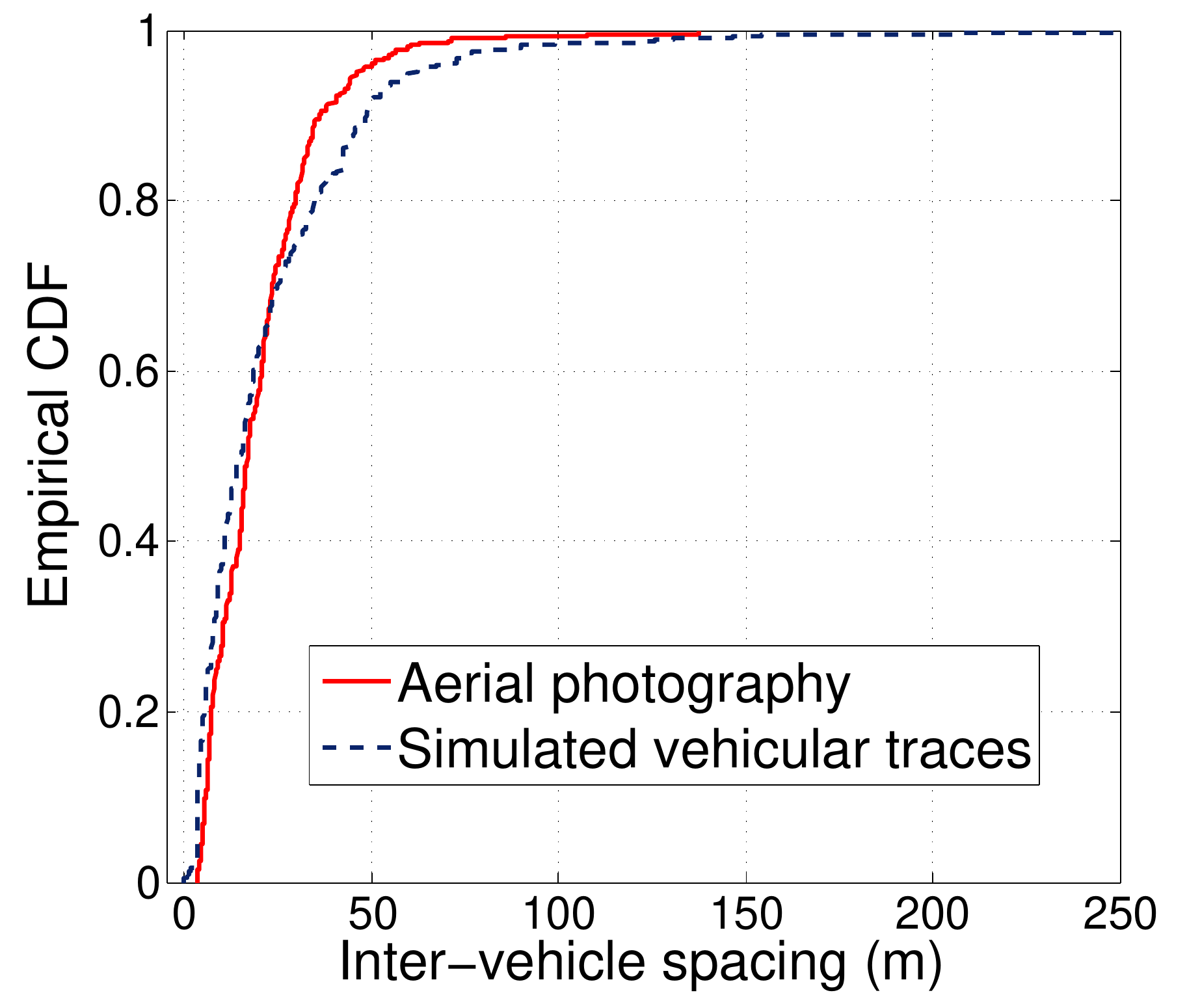}
     \caption{\small Inter-vehicle spacing for the simulated \emph{medium} density vehicular mobility trace and for the aerial photography of the A28 highway in Portugal.}
      \label{inter-veh-spacing}
   \end{center}
\end{figure}

In this section, we perform system-level simulations to evaluate the end-to-end, multi-hop performance of TVR and the two techniques discussed in Section~\ref{subsec:techniques}.  %
For this purpose, we generated vehicular traces using the STRAW mobility model~\cite{choffnes05} on a road of the same length (13.5~km), the same number of lanes (four), and similar shape to highway A28 where aerial imagery was acquired. We used three vehicular densities: 2.5, 7.5, and 10~vehicles/km/lane (designated in~\cite{naumov06} as \emph{low}, \emph{medium}, and \emph{high}, respectively) while keeping the same percentage of tall vehicles of approximately 14\% as observed in the aerial dataset. This resulted in 135, 404, and 675 vehicles in the system for different vehicular densities. The \emph{medium} density dataset was comparable to the A28 dataset (equal number of vehicles). We validated the traces against the aerial imagery by calculating the inter-vehicle distance (distance from each vehicle to its nearest neighbor for the generated \emph{medium} density and the A28 dataset). Figure~\ref{inter-veh-spacing} shows a good agreement between the cumulative distribution function of the inter-vehicle spacing for the generated \emph{medium} density traces and for the A28 highway, which also gives us confidence in drawing conclusions based on the generated vehicular traces for \emph{low} and \emph{high} densities. \\

In each generated vehicular mobility dataset (i.e., \emph{low}, \emph{medium}, and \emph{high}),
we randomly selected a set of source-destination pairs such that the source and destination are not direct neighbors. The number of analyzed source-destination pairs for each transmit power was 10000. To have a fair comparison, we used the same set of pairs to test all three techniques. %
The total number of source-destination pairs analyzed across different densities and transmit powers was $10^4\times3\times20=6\times10^5$. %

\begin{figure}
  \begin{center}
    \includegraphics[width=0.45\textwidth]{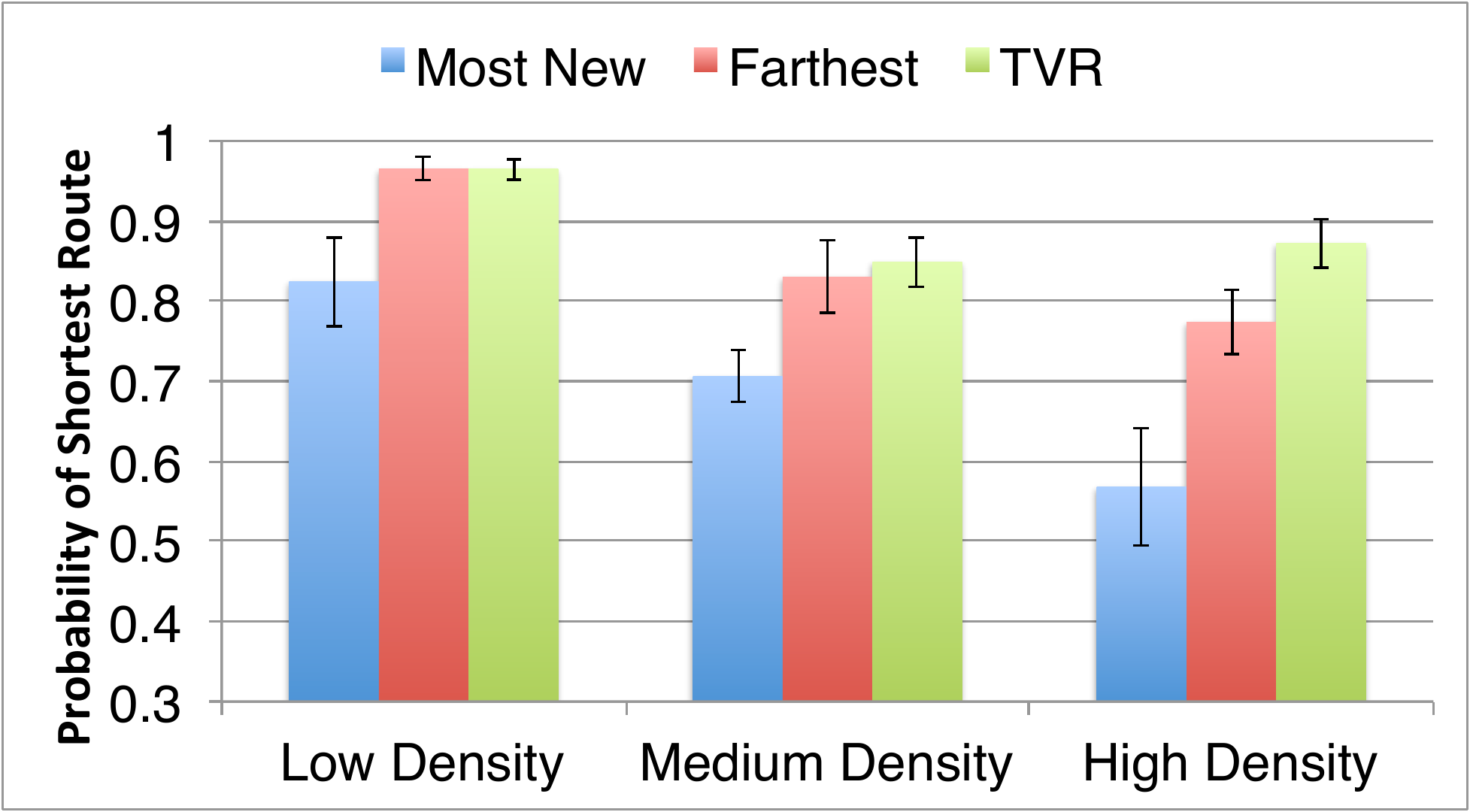}
     \caption{\small Performance of the three techniques in terms of the percentage of minimum hop routes from source to destination. Error bars represent one standard deviation drawn from the 20 different power settings (from 1 to 20 dBm).}
      \label{techniquesBestRouteComparison}
   \end{center}
\end{figure}

Figure~\ref{techniquesBestRouteComparison} shows the comparison of the three relaying techniques in terms of the probability of selecting a shortest (minimum-hop) route. Shortest route for a source-destination pair is defined as the least number of hops achieved by any of the three techniques. This was taken as a baseline: any of the techniques that had more than this number of hops did not choose the best route. %
Depending on the density and the employed technique, the average number of hops between the selected source and destination ranged between four and nine.

 TVR equals or outperforms the remaining two techniques, and as the density increases, its performance relative to the other two techniques improves. It is comparable to the \emph{Farthest Neighbor} technique at low density, on average 1.5 percentage points better than it at medium density, and 10 percentage points better at high densities. The reduced number of hops exhibited by TVR directly affects the end-to-end delay (fewer hops means a shorter time to get to the destination). %
 
It is interesting to see that the ratio of best routes per technique decreases as the vehicular density increases; this is due to the inability of any particular technique to always find %
 the best next relay. When the vehicular density is low, there are fewer neighbors to choose from, therefore choosing the one with best properties is easier. As the density increases, the ability to choose that specific relay decreases. \\

\subsection{Properties of Selected Best Hop Links}
\begin{figure}
  \begin{center}
    \includegraphics[width=0.35\textwidth]{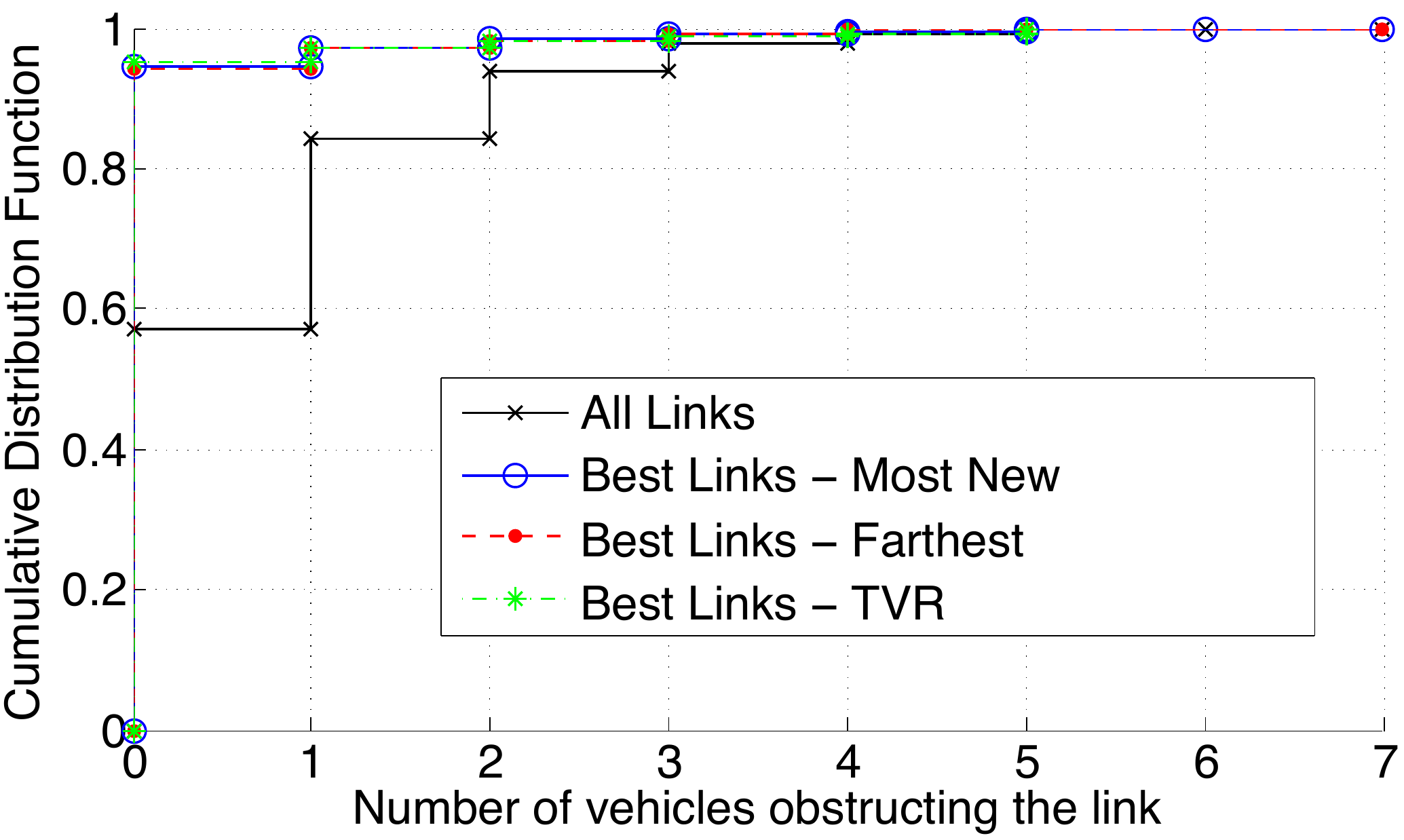}
     \caption{\small Difference between the number of obstructing vehicles in all links above the reception threshold in the system and the best links selected by the three employed techniques (\emph{Most New Neighbors}, \emph{Farthest Neighbor}, TVR). Tested on the aerial photography data of the A28 highway. Power settings: Tx Power 10~dBm; Receiver sensitivity threshold: -90~dBm. Other power settings exhibit similar behavior. %
}
      \label{allSchemesVsTall}
   \end{center}
\end{figure}

Figure~\ref{allSchemesVsTall} shows the number of vehicles obstructing the LOS for the links selected by the three techniques as well as all the links in the system. While system-wide only 58\% of links have LOS (i.e., zero obstructing vehicles), all three employed techniques select LOS links more than 92\% of the time. This result suggests that, apart from the distance of the relay, the LOS conditions of a link are important. All three techniques are implicitly preferring the LOS links: the next hop in the \emph{Most New Neighbors} technique will often have the most new neighbors due to privileged LOS conditions; with \emph{Farthest Neighbor} technique the farthest neighbor is most often that which has a LOS, therefore receiving the message above the threshold at farther distance; and TVR benefits from the height to reduce the chance of NLOS.

\subsection{How Often is a Tall Vehicle Relay Available?}

The measurements %
described in~\cite{boban11_3,wisit07,bai09} %
show that the inter-vehicle spacing for free-flow traffic
follows an exponential distribution. 
For a certain ratio $\gamma$ of tall vehicles ($0\leq\gamma\leq1$), %
the inter-vehicle spacing distribution for tall vehicles equals $f_{K}(k)=\gamma\lambda_s e^{-\gamma\lambda_s k}$,
where $\lambda_s$ is the inverse of the average inter-vehicle spacing %
in meters.
The probability $P_{T}$ of there being at least one tall vehicle relay within a certain average communication range $R$ is the complement of the probability of having zero tall vehicles within R. Therefore, $P_{T}=F_K(R) = 1- e^{-\gamma\lambda_s R}$, %
where $F_K(\cdot)$ is the Cumulative Distribution Function (CDF) of the inter-vehicle spacing between tall vehicles. It has to be noted that, in real situations, $R$ is going to be a variable that is dependent on many factors (transmission power, road surroundings, etc.), including the vehicle density, since the increased vehicular density will decrease the transmission range, as shown in~\cite{meireles10}. Therefore, we consider $R$ as an average communication range for which the value can be determined from experimental data such as that in Fig.~\ref{fig:effective-comm-range}.
However, for the employed TVR technique we are not interested in the existence of a tall vehicle within the entire $R$; rather, we are interested in a distance interval $[R-x_{max}, R]$, where $x_{max}$ is calculated %
as described in eq.~\ref{eq:Qfn} %
and $x_{max}\leq R$. Therefore, we have the following probability %
of having at least one tall vehicle relay within $[R-x_{max}, R]$:

\begin{align}\label{eqrkno}
P_{T[R-x_{max},R]} & =P_{T[0,x_{max}]} \\
\nonumber& = 1-Pr(k> x_{max}) \\
\nonumber& = F_K(x_{max}) \\
\nonumber& = 1-e^{-\gamma\lambda_s x_{max}}, 
\end{align} 
where the first step is a consequence of the memoryless property of the exponential distribution.
We analyze a fully connected network (i.e., at a certain point in time, each node has a route to all other nodes) with free-flow traffic\footnote{Free-flow traffic is defined as traffic where each vehicle is free to move at the desired speed~\cite{may90}, meaning the traffic volume is low enough so there are no traffic-induced decelerations. The converse of free-flow is high volume traffic near or in congestion. Arguably, in such a network, for the same ratio of tall vehicles, the probability distribution of tall vehicles, and therefore the probability of having a tall vehicle neighbor will be lower-bounded by %
eq.~\ref{eqrkno}.}. %

\begin{figure}
  \begin{center}
    \includegraphics[width=0.35\textwidth]{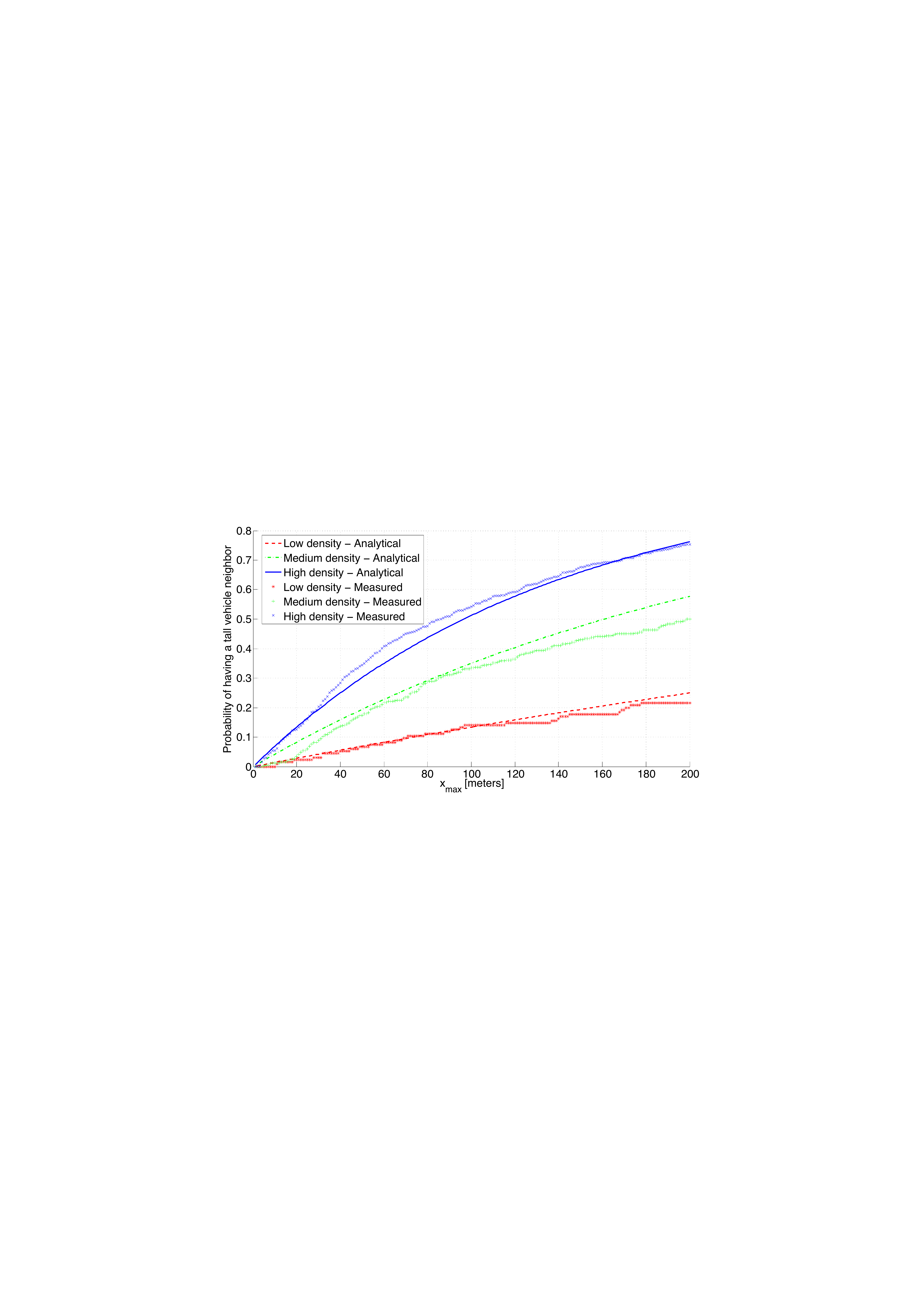}
     \caption{\small Probability of having a tall vehicle neighbor within $[R - x_{max}, R]$.}
      \label{ProbTallVehNeighbor}
   \end{center}
\end{figure}

Figure~\ref{ProbTallVehNeighbor} shows the analytical probability (eq.~\ref{eqrkno}) of having a tall vehicle neighbor within $[R - x_{max}, R]$ compared to that measured using aerial photography (medium density) %
and the generated vehicular traces (low and high density). %
There %
is a good match between the measured and analytical results; in both cases, the probability is approximately 35\% and 20\% when  $x_{max}=50$~meters for high and medium density, respectively. Only for low density the probability is below 20\% even with $x_{max}=150$~meters. This result explains why at higher densities TVR performs better: the increase in the overall number of neighboring vehicles increases the probability of having a tall vehicle within the $[R - x_{max}, R]$ region, thus enabling the selection of shorter routes via tall vehicles. In low density scenarios, there simply are not enough tall vehicles to make a positive difference, therefore TVR most often falls back to farthest neighbor relaying.

\begin{table}
	\centering
		\caption{\small Percentage of Vehicles Used for Relaying} 
\begin{tabular}{|c c c c|} \hline
		 \bf \hfill Density & \bf Low & \bf Medium & \bf High \\ 
		 \bf Technique &&& \\ \hline \hline
		\bf Most New Neighbors & 44\% & 40\% & 31\%\\ \hline
		\bf Farthest  & 34\% & 28\%&  26\%\\ \hline
		\bf TVR & 33\% & 27\% & 21\%\\ \hline
		\end{tabular} 
\label{tab:percentageRelaying}
\end{table}

\subsection{Does TVR Create Bottlenecks on Tall Vehicles?}
In this study, we focused on the effects of tall vehicle relaying in terms of per-hop increase in the received power (i.e., ``physical layer'') and improvement in end-to-end relaying by reducing the number of hops (i.e., ``network layer''), thus directly decreasing both the delay and the overall number of messages that need to be exchanged in the system (which in turn also reduces congestion). For both of these metrics, TVR was shown to perform better than other techniques. 
However, in our simulations, we assumed a perfect medium access scheme, which does not incur any contention or interference-induced losses.  %
Therefore, one question arises: if the majority of data traffic is relayed over tall vehicles, does this create bottlenecks -- situations where the tall vehicles cannot support the traffic being relayed over them? 
To answer this question, we analyzed the percentage of vehicles that are used for relaying as follows. For each technique, the same set of 10000 source-destination routes per vehicle density were taken into consideration, and the percentage of total number of vehicles used as relays by \emph{any} route has been reported in Table~\ref{tab:percentageRelaying} (results rounded to the nearest percentage point). As can be seen, the TVR technique does use a smaller percentage of vehicles; however, the difference is at most five percentage points %
when compared to the \emph{Farthest Neighbor} technique. %
Furthermore, this result also implies that neither of the techniques uses all vehicles in the system; rather, those vehicles are selected that have strategically better positions for relaying (e.g., a vehicle connecting two otherwise disconnected clusters, a vehicle that has a clear LOS with the most neighbors, etc). %

With respect to possible increase in the delay spread (and the resulting impact on the coherence bandwidth) due to relaying over tall vehicles, Paier et al. in ~\cite{paier07} as well as Acosta and Ingram in ~\cite{acosta06_2} performed experiments with tall vehicles (vans with heights comparable to those we used). The conclusion of both studies was that the maximum excess delay is mostly contained within 1 microsecond across measurement scenarios. Since the guard interval for IEEE 802.11p radios is set to 1.6 microseconds~\cite{dsrc09}, the IEEE 802.11p radios are capable of supporting the delay spread generated by the antennas mounted on tall vehicles.

By increasing the effective transmission range, TVR also increases the interference range. Therefore, similar to any relaying scheme that focuses on reducing the number of hops, TVR sacrifices the spatial reuse, since the increase in effective per-hop transmission range causes the increase in interference. The transmit power control can be employed in cases when it is necessary to increase the the spatial reuse, while retaining a more uniform coverage provided by the TVR. 

\section{Related Work} \label{sec:RelatedWork}

A number of VANET studies have pointed out the importance of antenna height in different contexts. %
The benefits of vertical antenna diversity were explored by Oh \emph{et al.} in~\cite{oh09}, where antennas were vertically displaced by 0.4~meters on a passenger car (i.e., a short vehicle) by installing one antenna inside the passenger cabin and a number of antennas on the car's roof. %
Both parking lot and on-road experiments were conducted using IEEE 802.11a radios operating in the 5.2~GHz frequency band. While mainly focusing on mitigating the negative effects of ground reflections rather than dealing with vehicular obstructions, the results show that the vertical diversity increases the effective communication range by more than 100 meters in certain scenarios. Kaul \emph{et al.} 
reported a similar study in~\cite{kaul07}, with a focus on determining the single best location for an antenna in a passenger car. By performing parking lot and on-road experiments using IEEE 802.11a radios operating in the 5.2~GHz frequency band, the center of the roof was found to be the best overall position, with significant variation in reception patterns when the antenna was displaced horizontally and vertically. On the other hand, two simulation studies based on detailed ray-optical channel models (\cite{reichardt09} and \cite{kornek10}) indicate that antenna positions  other than those on the roof can be preferable in certain scenarios (e.g., on side mirrors). %

With respect to Vehicle-to-Infrastructure (V2I) links and the impact of antenna placement, %
Paier \emph{et al.} in \cite{paier10_2} performed experiments %
which showed significantly better results %
with a road-side unit (RSU) that was placed above the height of the tallest vehicles. %
Placing the RSUs higher up %
results in a more reliable communication channel, which is particularly important for safety related applications. Since the RSU radio design is similar to the on-board unit (OBU) radios in vehicles, this finding suggests that the same applies for V2V communication; i.e., placing the antennas on taller vehicles is likely to result in improved radio channel. A similar study was reported in \cite{paier10}, where the authors analyzed the performance of a downlink between an RSU and an OBU installed in a vehicle. Antenna heights and traffic had a severe impact on the downlink performance, and the authors pointed out that ``shadowing effects caused by trucks lead to a strongly fluctuating transmission performance, particularly for settings with long packet lengths and higher speeds.'' This reinforces the findings reported by Meireles \emph{et al.} in \cite{meireles10}, where high losses were observed when obstructing vehicles were present between communicating vehicles.

Regarding the performance analysis and modeling of LOS and non-LOS (NLOS) channels, %
Tan \emph{et al.} \cite{tan08} performed V2V and V2I measurements in urban, rural, and highway environments at 5.9~GHz. The results point out significant differences with respect to delay spread and Doppler shift in case of LOS and NLOS  channels (NLOS was often induced by trucks obstructing the LOS). The paper distinguishes LOS and NLOS communication scenarios by coarsely dividing the overall obstruction levels. Similarly, Otto \emph{et al.}  \cite{Otto2009} performed V2V experiments in the 2.4~GHz frequency band in an open road environment and reported a significantly worse signal reception during a heavy traffic, rush hour period in comparison to a no traffic, late night period. In the WINNER project~\cite{baum05}, a series of 5.3 GHz wireless experiments were performed with a stationary base station and a moving node. The results were then used to derive channel models for use in simulation. Higher antenna heights were found to be beneficial to communication: the higher the antenna, the lower the path-loss exponent.
Several other experimental studies and surveys either discuss potential impact of vehicles %
on the channel quality:~\cite{abbas13,dhoutaut06,matolak08, matolak09}. %

Numerous relay selection metrics have been proposed for vehicular networks. The most common are: \textbf{1)} hop-count metrics (e.g., \cite{namboodiri04}); \textbf{2)} received power metrics (e.g., \cite{naumov06}); \textbf{3)} metrics based on geographic characteristics such as vehicle position, direction, or map information, etc. (e.g., \cite{naumov07}, \cite{lochert05}); and \textbf{4)} vehicular density based metrics (e.g., \cite{wisit07}). Combination of two or more of these metrics is also common in the literature. 
In this paper we have shown that relaying messages over tall vehicles is beneficial in terms of the hop count metrics (TVR results in fewer hops, particularly in dense vehicular networks) and received power metrics (tall vehicles exhibit higher received power, PDR, and communication range). %
Apart from our preliminary study reported in~\cite{boban11_2}, to the best of our knowledge, none of the existing studies proposed utilizing the %
information about the type and height of vehicles to improve the performance of V2V communication. %

\section{Conclusions} \label{sec:Discussion} %
We have determined the benefits of utilizing the height of vehicles %
to enable more efficient V2V communication.
We have shown that using knowledge about vehicle type/height to appropriately select the next hop vehicle consistently results in increased effective communication range and larger per-hop message reachability. %
Through both experiments and simulations that use a validated model, we have shown that tall vehicles are significantly better relay candidates than short vehicles when tall vehicles are within a certain distance of the farthest vehicle. Selecting tall vehicles in such situations results in a higher received signal power, %
increased packet delivery ratio, and larger effective communication range. 

Furthermore, we characterized the properties of preferred next hops in an experimental setting and by evaluating three relay techniques through system-level simulations. %
Both experiments and simulations showed that, when available, LOS links are preferred, regardless of the specific environment or relaying technique.
However,
since the distinction between LOS and NLOS links is not straightforward at the transmitter, %
we propose the tall vehicle relay (TVR) technique, which increases the likelihood of having a LOS link. %
We have shown that by selecting tall instead of farthest vehicles, %
TVR outperforms other techniques in terms of the number of hops to reach the destination, which in turn reduces end-to-end delay and congestion.  %
Therefore, the farthest neighbor metric might not be the best solution for selecting the next-hop relay where heterogeneous vehicle types exist (i.e., tall and short). The type of potential relay candidate can play an important role in deciding which next hop to select. %
Additionally, since TVR increases the received power level and reduces hop count, it can be used to improve performance of \emph{existing} routing protocols by adding binary information on the type of vehicle (tall or short).

It is important to note that our findings can be used to enhance different types of routing protocols, be it unicast~\cite{boban09}, broadcast~\cite{viriyasitavat11, tonguz10_2}, geocast~\cite{lochert05} or multicast~\cite{kihl07}. On highways, trucks and other tall commercial vehicles can be used as moving hotspots that relay the messages between the shorter vehicles. In urban environments, public transportation vehicles such as buses and streetcars can be used for the same purpose. %

\section*{Acknowledgements}
We are grateful to Carlos Pereira for his help during the experimental measurement and to Prof. Michel Ferreira for providing us with the aerial photography dataset of the A28 highway. We would also like to acknowledge Prof. Michel Ferreira and Dr. Tiago Vinhoza for participating in the initial discussions during which the problem was formed.
\bibliographystyle{IEEEtran}
\bibliography{draftIII_tex}

\end{document}